\documentclass[aps,nofootinbib,amsmath,prd,showpacs,groupedaddress,12pt,notitlepage,superscriptaddress]{revtex4-1}
\pdfoutput=1 

\usepackage[T1]{fontenc} 
\usepackage{tikz}
\usepackage{url}
\usepackage{amsfonts}
\usepackage{amsmath}
\usepackage{mathrsfs}
\usepackage[macrosonly]{chet}
\usepackage{hyperref}

\definecolor{darkblue}{rgb}{0.1,0.1,0.7}
\hypersetup{colorlinks,
           linkcolor={darkblue},
           citecolor={darkblue},
           urlcolor={darkblue}
}

\usetikzlibrary{shapes.misc}
\tikzset{cross/.style={cross out, draw=black, minimum size=2*(#1-\pgflinewidth), inner sep=0pt, outer sep=0pt},
cross/.default={1pt}}
\newcommand{\dm}{\mathrm{d}}

\newcommand{\de}{\partial}

\newcommand{\lsp}{\hspace{0.5pt}}

  {\left\lbrace\begin{array}{@{}l@{}}}%
  {\end{array}\right.}

\numberwithin{equation}{section}

\def\bea{\begin{eqnarray}}
\def\eea{\end{eqnarray}}
\def\be{\begin{equation}}
\def\ee{\end{equation}}
\def\ba{\begin{array}}
\def\ea{\end{array}}
\def\nn{\nonumber}

\def\nn{\nonumber}

\begin{document}


\title{Scale and Conformal Invariance \\ in Higher Derivative Shift Symmetric Theories}

\author{Mahmoud Safari}
\email{mahsafa@gmail.com}
\affiliation{Department of Computer Science, University of Freiburg, Georges-K\"ohler-Allee 074, 79110 Freiburg, Germany}

\author{Andreas Stergiou}
\email{andreas.stergiou@kcl.ac.uk}
\affiliation{Theoretical Division, MS B285, Los Alamos National
  Laboratory, Los Alamos, NM 87545, USA}
\affiliation{Department of Mathematics, King's College London, Strand, London WC2R 2LS, United Kingdom}

\author{Gian Paolo Vacca}
\email{vacca@bo.infn.it}
\affiliation{INFN - Sezione di Bologna, via Irnerio 46, I-40126 Bologna, Italy}

\author{Omar Zanusso}
\email{omar.zanusso@unipi.it}
\affiliation{Universit\`a di Pisa and INFN - Sezione di Pisa, Largo Bruno Pontecorvo 3, I-56127 Pisa, Italy}

\begin{abstract}
The critical behavior of infinite families of shift symmetric interacting
theories with higher derivative kinetic terms (non unitary) is considered.
Single scalar theories with shift symmetry are classified according to
their upper critical dimensions and studied at the leading non trivial
order in perturbation theory. For two infinite families, one with quartic
and one with cubic interactions, beta functions, criticality conditions and
universal anomalous dimensions are computed. At the order considered, the
cubic theories enjoy a one loop non renormalization of the vertex, so that
the beta function depends non trivially only on the anomalous dimension.
The trace of the energy momentum tensor is also investigated and it is
shown that these two families of QFTs are conformally invariant at the
fixed point of the RG flow.
\end{abstract}

\maketitle


\renewcommand{\thefootnote}{\arabic{footnote}}
\setcounter{footnote}{0}


\section{Introduction} \label{Section_introduction}

Quantum field theory (QFT) is a powerful framework with an illustrious
track record in modern theoretical physics. Applications to high-energy
physics typically require conditions like unitarity, locality, canonical
kinetic terms etc.\ to be imposed due to fundamental physical requirements.
There are physical systems, however, where such conditions can be relaxed
and in which the usual methods of QFT still produce relevant and important
results, e.g.\ \cite{Fisher:1978pf, Cardy:1985yy}, \cite{Landau:1986aog,
Riva:2005gd} and \cite{David:1992vv, David:1993rr}.

In this work our goal is to study the critical behavior of some shift
symmetric single field scalar theories in the perturbative $\varepsilon$
expansion below a critical dimension $d_c$. Our theories have higher
derivative kinetic terms and are thus non unitary, but are here viewed as
laboratories for testing standard renormalization group (RG) ideas in a new
setting.  We shall not try here to go beyond the scope of perturbation
theory to test the qualitative idea that well below the critical dimensions
--- in the non perturbative domain of the analysis, and in particular in
$d=4$ --- some of these critical theories survive. This is indeed a
difficult task, especially for non unitary theories.

There are many considerations that motivate this work: we would like to
develop a better understanding of the theory space of a single scalar field
in the non unitary sector, classify higher derivative shift symmetric
interactions, provide some explicit examples of interacting critical
theories of this kind, and consider the famous issue of scale vs conformal
invariance in this new arena. We consider in detail two families of
theories with a kinetic term of the form $\phi(-\square)^k\phi$ and even or
odd derivative interactions in the scalar field $\phi$, namely
$(\partial^\mu\phi\lsp \partial_\mu\phi)^2$ and
$\partial^\mu\phi\lsp\partial_\mu\phi\lsp \square\phi$, respectively. The
case $k=1$ corresponds to a canonical kinetic term, but we focus
generically on $k\ge2$.

We analyze the quantum criticality condition in the $\varepsilon$ expansion
below appropriate upper critical dimensions and study the universal
anomalous dimension of the field and, for the even family, also the scaling
of the shift symmetric quadratic in $\phi$ relevant operator
$\partial^\mu\phi \lsp \partial_\mu\phi$. Due to the derivative
interactions, our computations of RG quantities proceed naturally in position
space and not only in the more traditional momentum space. The structure of
the beta function in both classes of theories, quartic and cubic, is such
that fixed points of the RG are found within the $\varepsilon$ expansion
for each $k\geq 2$ and $k\geq 3$, respectively.  We thus find two infinite
families of new fixed points. For the even family the fixed points sit at
real values of the coupling for every $k$, but for the odd family the fixed
points sit at real (resp.\ imaginary) couplings for $k$ odd (resp.\ even).

We also analyze the issue of scale vs conformal
invariance~\cite{Polchinski:1987dy, Nakayama:2013is}. In shift symmetric
theories the Noether current for shift symmetry corresponds to a spin-one
operator whose divergence gives the equation of motion. Such a spin-one
operator is obviously conserved due to the equation of motion and cannot be
the virial current that is necessary in a theory that is scale but not
conformally invariant. Nevertheless, it is not clear that scale invariance
implies conformal invariance in our critical theories, especially because
of the fact that our kinetic term involves more than two derivatives which
renders our theories non unitary. As we will see by explicitly computing
the trace of the energy-momentum tensor, the fixed points that we find are
actually fully conformally invariant, at least at the investigated order of
perturbation theory, and so shift symmetry at this level is compatible with
conformality.

An interesting feature that we encounter in the odd case is that there are
no radiative vertex corrections and thus the beta function arises solely
due to wavefunction renormalization.
We emphasize that this result is obtained to the leading order of the $\varepsilon$ expansion.
Clearly, it deserves further investigation at higher orders.

The work is organized as follows: in section~\ref{classification} we
provide a general classification of the higher derivative shift symmetric
theories according to the order of the kinetic term and the form of the
interactions which are linked to the upper critical dimension.  Several
examples are considered, among which we identify the two simplest families
with even and odd interactions.  Section~\ref{Quartic_theories} is devoted
to the analysis of a family with quartic interaction, for which we employ
different techniques such as the standard perturbative renormalization
group, functional renormalization group~\cite{Wetterich:1992yh,
Morris:1994ie}, as well as the computation of the anomalous dimension using
Schwinger--Dyson equations.  Such techniques have previously been employed
in similar investigations of different kinds of higher derivative
theories~\cite{Safari:2017irw, Safari:2017tgs}.  In
section~\ref{Cubic_theories} we similarly analyze a family of odd
interacting theories, deriving the universal field anomalous dimensions and
the criticality conditions.  In section~\ref{EMT_CFT} we show that, for
both these families of critical theories, there exists an improved energy
momentum tensor with zero trace at the first non trivial order in
perturbation theory, implying that such scale invariant theories are also
conformal.  Finally, we present our conclusions and offer a few directions
for future work. Some computational aspects are summarized in two short
technical appendices.

\section{Shift symmetric theories with derivative interactions} \label{classification}

Let us classify possible realizations of shift symmetric single scalar field theories.
The action for a single scalar field $\phi$ must be invariant under the field transformation
\be
\phi(x) \to \phi(x) +c \,.
\label{shift}
\ee
Evidently, this is the case if every instance of the field appears under the action of a derivative operator.
Moreover, one clearly expects, for scalar theories, an even total number of
derivatives present in each operator because of Lorentz invariance (or
rotational invariance in Euclidean signature).
We shall be interested in two cases according to the structure of the
operators characterizing the interactions: (a) $\mathbb{Z}_2$ symmetric operators,
having an even order in the scalar fields and with interactions with at
least one derivative acting on each field, and (b)
operators with an odd order in the fields so that
we have the action of a double derivative on at least one of them.

Let us consider a theory characterized by the classically marginal kinetic
term with $2k$ derivatives, which, after a suitable integration by parts
which preserves the shift symmetry, can be written in a Minkowskian
signature with the mostly-plus metric as
\be\label{eq:kinetic-general}
- \tfrac{1}{2} \phi (-\Box)^k \phi \,,
\ee
and for which the canonical dimension of the scalar field is $\delta=\frac{d}{2}-k$.\footnote{%
For $k$ even the manifestly shift invariant kinetic term in the Lagrangian is
$-\frac{1}{2}(\Box^{\frac{k}{2}}\phi)^2$,
instead for $k$ odd it is
$-\frac{1}{2}(\partial_\mu \Box^{\frac{k'}{2}}\phi)^2$ where $k'=k-1$. Both reduce to \eqref{eq:kinetic-general} up to a boundary term.
}
We shall limit ourselves to the case of non-negative canonical dimension of
the field, $[\phi]\ge0$, i.e.\ $d_c \ge 2k$.

Considering a shift symmetric interaction given by a marginal operator with
$2l$ derivatives and $N$ fields of the form $ \partial^{2l} \phi^{N}$, where $N\le 2l$ to have all the fields under the action of at least one derivative,
one can easily derive a relation involving also the critical dimension $d_c$ and the order of the kinetic term $k$, as
\be
d_c=N \left( \frac{d_c}{2} -k \right)+2 l \quad  \Rightarrow \quad
d_c=2 \frac{\, k N-2l}{N-2}\quad \text{and} \quad \delta_c=2 \frac{k-l}{N-2}\,,
\label{basic_rel}
\ee
so that we generally require $k\ge l$ for a non negative field canonical dimension.

\subsection{\texorpdfstring{$\mathbb{Z}_2$}{Z\_2} symmetric theories} \label{Z2sect}

We start by asking which operators with an even number of fields, i.e.\ $N=2\alpha$, can play the role of  a marginal shift symmetric interaction of at least order $4$ in the scalar field at a given upper critical dimension $d_c$.
They are in general composite operators of the form $ \partial^{2l} \phi^{2\alpha}$, i.e.\ they contain $2l$ derivatives and $2 \alpha$ fields,
provided $ 2\le \alpha \le l $.

These inequalities for $\alpha$ can be rewritten using Eq.~\eqref{basic_rel} as
\be
2 \le \alpha=\frac{d_c\!-\!2l}{d_c\!-\!2k}\le l
\label{halffields}
\ee
and consequently one also deduces for $l$ the constraints
\be
\frac{d_c}{d_c+2\!-\!2k} \le l \le 2k-\frac{d_c}{2} \,.
\label{allowl}
\ee
Therefore, given the requirement $k\ge2$ which defines the free theory and independently from $l$,
one can solve the inequality for the critical dimension $d_c$ of any non trivial interaction and find
\be
2k \le d_c \le 4(k-1) \,.
\label{dcrange}
\ee

Let us now consider the possible theories allowed by some specific values of $k$.
We shall focus on the range $1\le k\le 4$, the extension to higher values being straighforward,
and then consider a specific family of theories characterized by a single quartic interaction with $l=2$.

\subsubsection*{\texorpdfstring{$k=1$}{k=1}}

This case is trivial and we include it for completeness.
Free fields of non negative dimension require $d_c\ge 2$. There is no room for shift symmetric interactions ($l \ge \alpha\ge2$).
Indeed, the inequalities in Eq.~\eqref{allowl} imply $l=1$.
So the only scale invariant and shift symmetric scalar theory of this kind is the free theory, which is obviously also conformal.

\subsubsection*{\texorpdfstring{$k=2$}{k=2}}

From Eq.~\eqref{dcrange}, we have $d_c=4$ and, from Eq.~\eqref{allowl}, one finds $l=2$ that also implies $\alpha=2$.
Therefore $(\partial \phi \cdot \partial \phi)^2$  is the only possible interacting marginal operator.
One then has a possible critical shift symmetric theory defined by
\be
-{\mathscr L}=\tfrac{1}{2}\phi \Box^2 \phi +g (\partial \phi \cdot \partial \phi)^2
\ee
for which the field $\phi$ is dimensionless at the critical dimension $4$.

\subsubsection*{\texorpdfstring{$k=3$}{k=3}}

From Eq.~\eqref{dcrange}, we have $6 \le d_c \le 8$.
Like the previous case, the operators describing non trivial marginal interactions have a number of derivatives that is bounded by Eq.~\eqref{allowl}.
One finds only two possible interacting non trivial general theories which possess shift symmetry,
either with ($d_c=6$, $l=3$) or with ($d_c=8$, $l=2$).

The most complex theory occurs for $d_c=6$ (massless field), whose Lagrangian can be written, for example, as
\bea
-{\mathscr L}=&{}&\tfrac{1}{2}\phi \left(-\Box\right)^3 \phi +g_1 (\partial \phi \cdot \partial \phi) \Box   (\partial \phi \cdot \partial \phi)+g_2 (\Box \phi )^2 (\partial \phi \cdot \partial \phi)\nonumber\\
&{}&+g_3 (\partial \phi \cdot \Box \partial \phi)  (\partial \phi \cdot \partial \phi) +h (\partial \phi \cdot \partial \phi)^3\,,
\eea
and the field is canonically massless.

For $d_c=8$ one has the much simpler theory
\be
-{\mathscr L}=\tfrac{1}{2}\phi \left(-\Box\right)^3 \phi +g (\partial \phi \cdot \partial \phi)^2\,,
\ee
where the field $\phi$ has canonical mass dimension $1$.

\subsubsection*{\texorpdfstring{$k=4$}{k=4}}

From Eq.~\eqref{dcrange} the possible theories must have $8 \le d_c \le 12$.
Once more, the operators describing non trivial marginal interactions have a number of derivatives bounded by Eq.~\eqref{allowl}.
One finds only four possible interacting non trivial general theories which possess shift symmetry, with $d_c=8, 9, 10, 12$ and $l=4, 3, 3, 2$, respectively.

If $d_c=8$, the field is canonically dimensionless and one finds several possible interactions described by operators with $8$ derivatives acting on $4$, $6$ and $8$ fields:
\bea
-{\mathscr L}=&{}&\tfrac{1}{2}\phi \left(-\Box\right)^4 \phi +\left( g_1 (\partial \phi \cdot \partial \phi) \Box^2   (\partial \phi \cdot \partial \phi)+\cdots\right)\nonumber\\
&{}&+ \left( h_1 (\partial \phi \cdot \partial \phi)^2 \Box  (\partial \phi \cdot \partial \phi)+\cdots \right)
+w (\partial \phi \cdot \partial \phi)^4 \,.
\eea

For $d_c=9$, the field has canonical mass dimension $1/2$. One finds a theory where only a marginal interaction with $6$ fields is allowed:
\be
-{\mathscr L}=\tfrac{1}{2}\phi \left(-\Box\right)^4 \phi +g (\partial \phi \cdot \partial \phi)^3\,.
\ee

For $d_c=10$, the field has unit canonical mass dimension. One finds a theory with three marginal interactions with $4$ fields and $6$ derivatives
\be
-{\mathscr L}=\tfrac{1}{2}\phi \left(-\Box\right)^3 \phi + g_1 (\partial \phi \cdot \partial \phi) \Box   (\partial \phi \cdot \partial \phi)+\cdots \,.
\ee

Finally, for $d_c=12$, the field has canonical mass dimension $2$ and one finds a theory where only a marginal interaction with $4$ fields is allowed:
\be
-{\mathscr L}=\tfrac{1}{2}\phi \left(-\Box\right)^4 \phi +g (\partial \phi \cdot \partial \phi)^2\,.
\ee

\subsubsection*{\texorpdfstring{$k \ge 5$}{k>=5}}

Starting from $k\ge 5$ also shift symmetric theories with fractional upper critical dimensions are allowed.  Indeed, for $k=5$, one finds the possible
upper critical dimensions $d_c=10, 32/3, 11, 12, 14, 16$ where the case with $d_c=32/3$ appears for $k=5$, $l=4$ and $\alpha=4$.
For higher values of $k$ many more fractional values of $d_c$ appear.

\subsubsection*{\texorpdfstring{$l=2$}{l=2}}

All the critical theories with $k\ge2$ and the single quartic interaction
corresponding to $l=2$ have a critical dimension that saturates the upper bound of the
inequality of Eq.~\eqref{dcrange}, i.e.\ $d_c=4(k-1)=4,8,12, \ldots,$ with general Lagrangian
\be\label{eq:general-family-quartic}
-{\mathscr L}=\tfrac{1}{2}\phi \left(-\Box\right)^k \phi +g (\partial \phi \cdot \partial \phi)^2\,,
\ee
which covers the $l=2$ special cases covered above.
In this family of models the lowest value $k=2$ is special since then the scalar field is canonically dimensionless.

\subsection{Theories with \texorpdfstring{$\mathbb{Z}_2$}{Z\_2} odd operators} \label{NonZ2sect}

Other interesting theories with shift symmetric interactions can be characterized by the presence of operators with an odd number of fields.
Depending on the dimension $d_c$ there can also be even interactions.
We shall focus here on the case of purely odd interactions.

Proceeding in a similar way to the even case, one considers the number of fields $N=2n+1$ with $n$ constrained to have values in the range $1\le n \le l-1$.
Using Eq.~\eqref{basic_rel} one finds
\be
3 \le (2n+1)=2 \frac{d_c\!-\!2l}{d_c\!-\!2k} \le 2l -1 \,.
\label{halffields2}
\ee
Therefore, at fixed $k$ and $d_c$, one obtains for $l$ the inequalities
\be
\frac{\frac{3}{2} d_c-k}{d_c-2k+2} \le l \le \frac{3}{2} k-\frac{d_c}{4} \,,
\label{allowlodd}
\ee
which guarantee that the corresponding operators are shift symmetric.

Again, defining the free theory for a given $k\ge2$ independently from $l$,
and solving the inequality, one finds the range allowed for the upper critical dimension $d_c$
\be
2k \le d_c \le 2(3k-4) \,.
\label{dcrangeodd}
\ee

For a given $d_c$, there can be theories with both kinds of marginal operators (even and odd number of fields), clearly with a different number of derivatives,
provided that $d_c$ is in the range that allows for the existence of both kinds of operators.
Conversely, one can find families of critical theories with only odd interactions.
This is the case when the following inequalities\footnote{%
A special case
for $d_c=6$ and $k=3$, satisfying Eq.~\eqref{dcrangeodd}, can be obtained
from the Lee--Yang model with field $\tilde\phi$, considering a change of field variable $\tilde \phi=\Box\phi$,
which makes the original field $\tilde \phi$ no longer a primary operator in the sense of CFT.
The corresponding theory enjoys the action $\int {\rm d}^dx \left(\frac{1}{2} \phi (-\Box)^3 \phi +\frac{g}{3!} (\Box \phi)^3\right)$.
Note that, since the field is canonically dimensionless, one can further deform the theory by adding marginal shift symmetric operators with up to six fields.
}
are satisfied:
\be
4(k-1)<d_c\le 2(3k-4), \quad k > 2.
\ee
For example, at the value of the upper bound of the critical dimension,
$d_c=2(3k-4)$ the only allowed interaction is characterized by $l=2$ and
$(2n+1)=3$ for any $k>2$, i.e.\ a cubic interaction with four derivatives,
which corresponds to the following family of critical theories with
$d_c=10,16,22,\ldots$:
\be
-{\mathscr L}=\tfrac{1}{2}\phi \left(-\Box\right)^k \phi +g\, \partial \phi \cdot \partial \phi \, \Box \phi
 \,.
\ee
We stress that the one above is the only possible independent shift symmetric cubic interaction with four derivatives, since the other possible cubic scalar operator,
$\partial_\mu\partial_\nu \phi \partial_\mu \phi \partial_\nu \phi$, differs from the one above by a total divergence term.
In this case, the scalar field at the upper critical dimension $d_c$ has a canonical dimension $\delta_c=\frac{d_c}{2}-k=2(k-2)$.

We are not going to dive any further into a systematic general classification based on the relations above.
As for the last examples that we want to convey, one can easily see that the first case of a critical theory with purely quintic interaction
$N=2n+1=5$ with integer dimension $d_c=16$ ($\delta_c=2$) appears for $k=6$ and $l=3$.
In the case of odd theories, fractional upper critical dimensions $d_c$ do appear for $k \ge 4$. For example, if $k=4$ one finds, in general,
the possible values $d_c=8, 28/3, 12,16$ and $d_c=28/3$, the latter corresponding to $l=3$ and $N=2n+1=5$.
For even higher values of $k$, there are several quintic theories with fractional $d_c$.
All these specific quintic theories have a Lagrangian of the form
\be
-{\mathscr L}=\tfrac{1}{2}\phi \left(-\Box\right)^k \phi +g\, (\partial \phi \cdot \partial \phi)^2 \, \Box \phi \,.
\label{quintic}
\ee

\section{Quartic models: \texorpdfstring{$\boldsymbol{\varepsilon}$}{epsilon}-expansion analysis} \label{Quartic_theories}

We shall focus on a perturbative analysis of the following family of
theories for $k\ge2$, which in the mostly-plus Minkowski signature is described by the actions
\be
S_M=- \int {\rm d}^{d}x \left[ \frac{1}{2}\phi \left(-\Box\right)^k \phi +\frac{g}{4} (\partial \phi \cdot \partial \phi)^2\right]\,,
\label{theories4_Minkowsky}
\ee
below the upper critical dimensions $d_c=4(k-1)$ at $d=d_c-\varepsilon$, discussed above in Eq.~\eqref{eq:general-family-quartic}.

It is convenient to perform the Wick's rotation from the very beginning using $x^0=-i x_4$, so that one has the well known relation between the Minkowskian and Euclidean actions, $i S_M=-S_E$, and the functional integral can be written in term of the Euclidean action $S=S_E$
\be
S= \int {\rm d}^{d}x \left[ \frac{1}{2}\phi \left(-\Box\right)^k \phi +\frac{g}{4} (\partial \phi \cdot \partial \phi)^2\right]\,.
\label{theories4}
\ee

The Green function (i.e.\ the propagator) of the massless free theory ($g=0$) satisfies, in the
coordinate representation, the differential equation
\be
  (-\square_x)^k G = \delta_x \,,
\ee
where we adopt the notation $\delta_x=\delta^{(d)}(x)$.
Choosing the normalization corresponding to the $1/(p^2)^k$ propagator in momentum space, by Fourier transforming one finds
\be \label{G}
G(x) = \frac{c}{|x|^{2\delta}}, \qquad c \equiv \frac{1}{(4\pi)^k\Gamma(k)}\,\frac{\Gamma(\delta)}{\pi^{\delta}}, \qquad
\delta=\frac{d}{2}-k=k-2-\frac{\varepsilon}{2} \,.
\ee
The operator $(\partial\phi\cdot\partial\phi)^2$ is a relevant deformation
of the free theory since
$\Delta_{(\partial\phi\cdot\partial\phi)^2}=d-\varepsilon<d$ when $g=0$.

\subsection{One loop perturbative $\texorpdfstring{\beta}{beta}$ function}

\begin{figure}[t]
        \scalebox{2}{
        \begin{tikzpicture}[baseline=-.1cm]
        \draw (0,0) circle (.5cm);
        \filldraw [gray!100] (.5,0) circle (2pt);
        \draw (.5,0) circle (2pt);
        \filldraw [gray!100] (-.5,0) circle (2pt);
        \draw (-.5,0) circle (2pt);
        \end{tikzpicture}
        }
        \caption{Diagram in coordinate representation for the one loop corrections to the quartic operator.}
        \label{CTint1loop4}
 \end{figure}
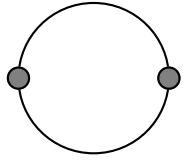

We compute here the leading order perturbative beta function of the coupling $g$ in the $\overline{\mathrm{MS}}$ scheme
from the one loop counterterm with the $1/\varepsilon$ pole, corresponding to
a Feynman diagram in which each one of the two external vertices is attached
to two external lines in various internal configurations.
More precisely let us write the counterterm in the coordinate representation, depicted in Fig.~\ref{CTint1loop4}, for the interaction given in Eq.~\eqref{theories4}, picking up the divergent part of the second order corrections
\be
S^{\rm CT}_{\rm int,2}=\int \prod_{i=1}^4 \dm^d x_i \left[ \frac{1}{2} \frac{1}{2^2} \frac{\delta^{2} S_{\rm int}}{\delta \phi(x_1) \delta \phi(x_2)} ( G_{x_{13}} G_{x_{24}}+G_{x_{14}} G_{x_{23}} )
\frac{\delta^{2} S_{\rm int}}{\delta \phi(x_3) \delta \phi(x_4)}\right]\,,
\label{generic_bubble}
\ee
where $x_{ij}=x_i-x_j$ and with
\be
\frac{\delta^{2} S_{\rm int}}{\delta \phi(x_1) \delta \phi(x_2)}= g \int \dm^d x_0 \, \de_\mu \delta_{x_{01} }
\left( \de_\nu \phi \de_\nu \phi\,  \de_\mu \delta_{x_{02}} +2 \de_\mu \phi \de_\nu \phi \de_\nu \delta_{x_{02}}\right)
\ee
which is obtained from the interacting part of the action. One finds
\bea
S^{\rm CT}_{\rm int,2}= \frac{g^2}{4} \int \dm^d x \dm^d y \Bigl\{&{}&
(\de \phi)_x^2 (\de_\mu \de_\nu G)_{xy}^2 (\de \phi)_y^2
+4 (\de \phi)_x^2 (\de_\mu \de_\nu G)_{xy} (\de_\mu \de_\beta G)_{xy} (\de_\nu \phi \de_\beta \phi)_y\nonumber \\
&{}&+  4  (\de_\mu \phi \de_\alpha \phi)_x (\de_\mu \de_\nu G)_{xy} (\de_\alpha \de_\beta G)_{xy} (\de_\nu \phi \de_\beta \phi)_y
\Bigr\} \,,
\eea
where the subscript ``$xy$'' in a Green function or in its derivatives stands for the translational invariant dependence in $x\!-\!y$.

In the integrand above, the divergent part with a $1/\varepsilon$ pole is proportional to a $\delta_{xy}$ as expected from the locality of the counterterm.
To extract it, it is convenient to move to the momentum representation as
intermediate step. Using some of the relations given in appendix~\ref{abeta}, one obtains
\bea
(\partial_\mu\phi\partial_\nu\phi\partial_\rho\phi\partial_\sigma\phi )\;  \partial_\mu\partial_\nu G\, \partial_\rho\partial_\sigma G &=& \frac{3}{2} \frac{1}{(4\pi)^{2k-2}\Gamma (2k)}\frac{1}{\varepsilon}\, (\partial\phi \cdot \partial\phi )^2 \delta_{xy} + \mathrm{finite} \\
((\partial\phi \cdot \partial\phi )\, \partial_\nu\phi\partial_\sigma\phi)\;  \partial_\rho\partial_\nu G\, \partial_\rho\partial_\sigma G &=& \frac{1}{2(4\pi)^{2k-2}\Gamma (2k)}\frac{1}{\varepsilon} (d_c+2)  (\partial\phi \cdot \partial\phi )^2  \delta_{xy} + \mathrm{finite}\nonumber \\
(\partial\phi \cdot \partial\phi )^2 \;(\partial_\mu\partial_\nu G)^2 &=& \frac{1}{2(4\pi)^{2k-2}\Gamma (2k)}\frac{1}{\varepsilon} d_c(d_c+2)  (\partial\phi \cdot \partial\phi )^2 \delta_{xy} + \mathrm{finite} \nonumber
\eea
so that, as expected, the one loop counterterm to the action contains a divergent local interaction which renormalizes the coupling $g$
\be
S^{\rm CT}_{\rm int,2}=  \frac{2(4k^2-2k+3)}{(4\pi)^{2k-2}\Gamma (2k)}\frac{1}{\varepsilon}\frac{g^2}{4} \int \dm^d x(\partial\phi \cdot \partial\phi )^2+ \mathrm{finite}\,.
\label{vertexCT}
\ee
No further counterterm, including the wavefunction renormalization, is needed at this order of perturbation theory.
From the counterterm, we deduce the $\overline{\mathrm{MS}}$ beta function for the coupling of the theories with upper critical dimension $d_c = 4k-4$ and $k\ge 2$.
\be
\mu\frac{dg}{d\mu} = \beta_4(g)=-\varepsilon g +\frac{2(4k^2-2k+3)}{(4\pi)^{2k-2}\Gamma(2k)} g^2\,.
\label{beta4}
\ee
It is straightforward to see that $\beta_4(g)=0$ has an ${\cal O}(\varepsilon)$ fixed point. As a consequence, the non trivial fixed point of the RG flow corresponds to a family of critical theories with the LO coupling
\be
\label{gFP4}
g=\frac{(4\pi)^{2k-2} \Gamma(2k)}{2 (4k^2-2k+3)} \varepsilon\,.
\ee
This is an attractive fixed point since
$\Delta_{(\partial\phi\cdot\partial\phi)^2}=d+\partial_g\beta_4(g)|_{g~{\rm of}~\eqref{gFP4}}= d+\varepsilon > d$.

\subsubsection{Alternative derivation with IR regulated functional RG methods}

We show here how to derive the one loop beta function \eqref{beta4} for the coupling $g$ using the Wilsonian Functional RG (FRG) approach for the IR regulated 1PI generator, the so-called effective average action $\Gamma_\kappa[\phi]$, which flows according to the Wetterich-Morris equation~\cite{ Wetterich:1992yh,Morris:1994ie}
\be
\dot \Gamma_\kappa[\phi]=\frac{1}{2} {\rm Tr}\left[ \left(\Gamma_\kappa^{(2)}+R_\kappa\right)^{-1} \dot R_\kappa \right]\,,
\label{WMeq}
\ee
where $\kappa$ is the IR mass scale, $R_\kappa$ is an operator describing the adopted coarse-graining scheme, and the dots represent $\log \kappa$ derivatives.
The use of a FRG scheme may be appealing for future non perturbative investigations.

The leading order perturbative beta function can be computed using the simplest possible truncation
\be
\Gamma_\kappa=\int {\rm d}^{d}x \left[ \frac{1}{2}\phi \left(-\Box\right)^k \phi +\frac{g_\kappa}{4} (\partial \phi \cdot \partial \phi)^2\right] \,.
\label{truncation}
\ee
In the momentum representation, we choose the simplest cutoff profile,
$R_\kappa(q^2)=(\kappa^{2k}-(q^2)^k)\theta(\kappa^2-q^2)$.
In order to proceed further, we first compute the second functional derivative of the action with respect to $\phi(x)$ and $\phi(y)$
\be
\Gamma_{\kappa,xy}^{(2)}=  (-\Box)^k\delta_{xy}  +g_\kappa \left[ -(\partial_\mu \phi \partial_\mu \phi) \Box_x \delta_{xy}
-2(\Box \phi \partial_\mu \phi +2\partial_\mu\partial_\nu \phi \partial_\nu \phi)
\partial_\mu \delta_{xy} - 2 \partial_\mu \phi \partial_\nu \phi \partial_\mu\partial_\nu  \delta_{xy}
\right]
\ee
Because of the simple form of the one loop perturbative corrections to the interaction operator, which diagrammatically has just two vertices,
the functional trace can be computed at his order in a very simple way: first substituting
the representation of the Dirac delta distribution, $\delta_{xy}=\int \frac{{\rm d}^d q}{(2\pi)^d} e^{iq \cdot (x-y)}$, to get
\be
\hspace{-0.3cm} \tilde\Gamma_\kappa^{(2)}(x; q)=(q^2)^\kappa+  g_\kappa \left[ q^2 (\partial_\mu \phi \partial_\mu \phi)
-2 i q_\mu (\Box \phi \partial_\mu \phi  +2\partial_\mu\partial_\nu \phi \partial_\nu \phi)
 +2q_\mu q_\nu  \partial_\mu \phi \partial_\nu \phi
\right]
\ee
and then inserting it in the functional trace on the right-hand side of Eq.~\eqref{WMeq}, which now reads
\be
\frac{1}{2}\int \dm^d x \int \frac{{\rm d}^d q}{(2\pi)^d}\frac{2s \,\kappa^{2k} \theta(\kappa^2-q^2)}{\kappa^{2k}+
g_\kappa \left[ q^2 (\partial_\mu \phi \partial_\mu \phi)  -2 i q_\mu (\Box \phi \partial_\mu \phi
+2\partial_\mu\partial_\nu \phi \partial_\nu \phi) +2q_\mu q_\nu  \partial_\mu \phi \partial_\nu \phi \right]}\,.
\ee
The perturbative one loop flow of the coupling $g_\kappa$ can be extracted by expanding this expression to order $g_\kappa^2$
and truncating the right hand side to contributions of the form of the operator $(\partial \phi \cdot \partial \phi)^2$. One obtains,
recalling that $d=4(k-1)-\varepsilon$ and neglecting $\varepsilon$ at this order,
\be
\frac{1}{2} g_\kappa^2 \frac{d^2+6d+20}{(4\pi)^\frac{d}{2} \Gamma(\frac{d}{2}+2)} \frac{1}{4} \int \dm^d x (\partial \phi \cdot \partial \phi)^2=
2 \frac{4k^2-2k+3}{(4\pi)^{2k-2}\Gamma(2k)} g_\kappa^2 \frac{1}{4} \int \dm^d x (\partial \phi \cdot \partial \phi)^2\,.
\ee
Equation~\eqref{WMeq} accounts for the coarse-graining procedure of the Wilsonian RG paradigm, which should be combined with a rescaling procedure.
Rescaling the spacetime by $1/\kappa$ and the field $\phi$ by $\kappa^{d_\phi}$ where $d_\phi$ is the field dimension, one should work with the beta functional defined with the rescaling applied. In the new units,
\be
\dot \Gamma_\kappa[\phi]=(-d +\#_\partial+d_\phi \#_\phi) \Gamma_\kappa+
\frac{1}{2} {\rm Tr}\left[ \left(\Gamma_\kappa^{(2)}+R_\kappa\right)^{-1} \dot R_\kappa \right]\,,
\label{scaledWMeq}
\ee
where $\#_\partial$ and $\#_\phi$ count in an operator the number of derivatives and the number of fields, respectively.
Inspecting the coefficients for $\frac{1}{4}(\partial \phi \cdot \partial \phi)^2$, one then gets
\be
\beta_g=-\varepsilon g +2 \frac{4k^2-2k+3}{(4\pi)^{2k-2}\Gamma(2k)} g^2\,,
\ee
in agreement with the previous analysis.

Let us continue within this framework, namely using Eq.~\eqref{scaledWMeq}, by investigating another universal property of these theories, in particular the leading perturbative corrections to the scaling dimension of the most relevant quadratic shift invariant deformation given by the composite operator
\be
O_2=\tfrac{1}{2} \lsp\de \phi \cdot \de \phi \,,
\ee
controlled by the coupling $m_2$.

In order to compute the associated critical exponent, it is enough to add to $\Gamma_\kappa$ the term $\int \dm^dx\, m_{2\kappa} O_2$ so that $\tilde\Gamma_\kappa^{(2)}(x; q) \to \tilde\Gamma_\kappa^{(2)}(x; q) +m_{2\kappa} q^2$.
Let us write the mass dimension of $O_2$ as $d^{\rm}_{O_2}\!=2\!+\!2\delta\!+\!\gamma_{O_2}\!\simeq2(k\!-\!1)\!-\!\varepsilon+\gamma_{O_2}$ and, similarly, for the coupling $m_2$ introduce $d_{m_2}=d- d^{\rm}_{O_2}\!\simeq2(k-1)\!-\!\gamma_{O_2}$. Note that at this order one can neglect the field anomalous dimension $\gamma={\rm o} (\varepsilon)$, which we shall compute in next sections.
Expanding perturbatively at second order the denominator in the functional trace, as discussed in~\cite{Codello:2017hhh},
one can obtain the leading order universal coefficients.
In particular one finds for the rescaled $\tilde{m}_{2\kappa}=m_{2\kappa}/\kappa^{2(k-1)}$
the universal flow (which can be also derived in the $\overline{\mathrm{MS}}$ scheme)
\be
k \frac{d}{dk} \tilde{m}_{2\kappa}=-2(k-1) \tilde{m}_{2\kappa}+\frac{4k-2}{(4\pi)^{2k-2} \Gamma(2k-1)} g_\kappa \tilde{m}_{2\kappa} = - d_{m_2} \tilde{m}_{2\kappa}\, .
\ee
Substituting the $g(\varepsilon)$ relation \eqref{gFP4} valid at criticality, one obtains at leading order in $\varepsilon$ the universal scaling exponent
\be
 d_{m_2}=2(k-1) -\frac{(2k-1)^2}{4k^2-2k+3} \varepsilon\,, \quad
\tilde{\nu}=\frac{1}{d_{m_2}}=\frac{1}{2(k-1)}+\frac{(2k-1)^2}{4(k-1)^2 (4k^2-2k+3)} \varepsilon \,,
\ee
the latter being the analogous of the standard critical exponent $\nu$, but for the double derivative of the two point function.
Clearly the leading order anomalous dimension of the deforming operator is
\be
\gamma_{O_2}=\frac{(2k-1)^2}{4k^2-2k+3} \varepsilon\,,
\ee
with $\gamma_{O_2} \to \varepsilon$ for $k \to \infty$.
The scaling dimension $\gamma_{O_2}$ can equivalently be computed inserting ${O}_2$ as composite operator and using the $\overline{\mathrm{MS}}$ formalism of the previous section, but it was somehow more straightforward to obtain it with FRG methods as shown here.

\subsection{Anomalous dimension}

\begin{figure}[t]
     \scalebox{2}{
     \begin{tikzpicture}[baseline=-.1cm]
     \draw (0,0) circle (.5cm);
     \draw (-.5,0) to [out=0,in=180] (.5,0);
     \filldraw [gray!100] (-.5,0) circle (2pt);
     \draw (-.5,0) circle (2pt);
     \filldraw [gray!100] (.5,0) circle (2pt);
     \draw(.5,0) circle (2pt);
     \path(0,-0.8) node {\scalebox{0.5}{(a)}};
     \end{tikzpicture}
     }
     \hspace{3cm}
     \scalebox{2}{
     \begin{tikzpicture}[baseline=-.1cm]
     \draw (0,0) circle (.5cm);
     \draw (0,.5) to [out=-90,in=30] (-.469,-.171);
     \draw (0,.5) to [out=-90,in=150] (.469,-.171);
     \filldraw [gray!100] (0,.5) circle (2pt);
     \draw (0,.5) circle (2pt);
     \filldraw [gray!100] (.469,-.171) circle (2pt);
     \draw (.469,-.171) circle (2pt);
     \filldraw [gray!100] (-.469,-.171) circle (2pt);
     \draw (-.469,-.171) circle (2pt);
     \def\x{2}

     \draw (0+\x,0) circle (.5cm);
     \draw (-.5+\x,0) to [out=0,in=180] (.5+\x,0);
     \filldraw [gray!100] (-.5+\x,0) circle (2pt);
     \draw (-.5+\x,0) circle (2pt);
     \filldraw[fill=white](.5+\x,0) circle (2pt);
     \draw (.5+\x,0) node[cross=1.9pt,rotate=0, black]{};

     \path(0.5*\x,-0.8) node {\scalebox{0.5}{(b)}};

     \end{tikzpicture}
     }

        \caption{Diagrams in coordinate representation for the corrections
        to the kinetic operator: (a) The $k>2$ case receives the leading
      order contributions at two loops of order $g^2$; (b) the $k=2$ case
    gets leading order contributions at three loops of order $g^3$.}
        \label{CTkinLO4}
 \end{figure}
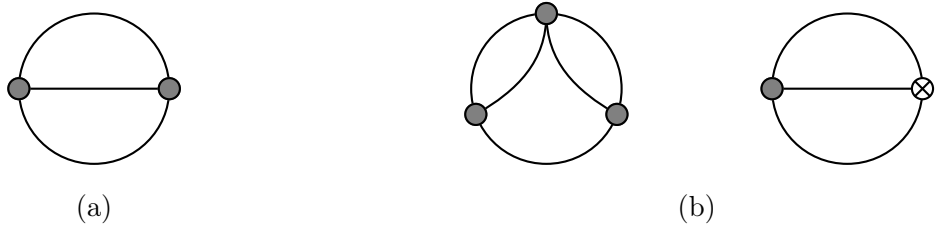

We include now the leading order contribution to the field's anomalous dimension.
We find that for $k>2$ the leading contribution appears at two loops as usual in theories with quartic interactions.  For $k=2$, where the canonical field dimension is zero, the anomalous dimensions gets its first contributions at three loops, instead.
In Fig.~\ref{CTkinLO4} we show the diagrams in the coordinate representation that give rise to divergent contributions in the form of the kinetic operator.

 \subsubsection{Leading order anomalous dimension for \texorpdfstring{$k>2$}{k>2}: perturbative RG approach}

The counterterm for the kinetic operator comes from the UV-divergent part of the following expression, as illustrated in Fig.~\ref{CTkinLO4}-(a),
\be
S^{\rm CT}_{\rm kin,2}=\int \prod_{i=1}^6 \dm^d x_i \left[ \frac{1}{2} \frac{1}{3!^2} \frac{\delta^{3} S_{\rm int}}{\delta \phi(x_1) \delta \phi(x_2)\delta \phi(x_3)} ( G_{x_{14}} G_{x_{25}} G_{x_{36}}+{\rm 5 \, perm.} )
\frac{\delta^{3} S_{\rm int}}{\delta \phi(x_4) \delta \phi(x_5)  \delta \phi(x_6)}\right] \,,
\label{prop2loopprop}
\ee
where
\begin{equation}
\begin{aligned}
\frac{\delta^{3} S_{\rm int}}{\delta \phi(x_1) \delta \phi(x_2)\delta
\phi(x_3)}= 2 g \!\! \int \dm^d x_0 \,\de_\nu \phi \big(
\de_\mu \delta_{x_{01} }  \de_\mu \delta_{x_{02} }  \de_\nu \delta_{x_{03}
} &+
\de_\mu \delta_{x_{01} }  \de_\mu \delta_{x_{03} }  \de_\nu \delta_{x_{02}
} \\&\qquad+
\de_\nu \delta_{x_{01} }  \de_\mu \delta_{x_{02} }  \de_\mu \delta_{x_{03} }
\big)\,.
\end{aligned}
\label{S3}
\end{equation}
Therefore, after some algebra, one finds
\be
S^{\rm CT}_{\rm kin,2}= -g^2 \int  \dm^d x  \dm^d y \,  (\de_\mu \phi)_x  \left(
(\de_\alpha \de_\beta G )^2  \de_\mu \de_\nu G + 2  \de_\mu \de_\beta G\, \de_\beta \de_\alpha G\, \de_\alpha \de_\nu G
\right)_{xy} (\de_\nu \phi)_y\,,
\label{kineticCT1}
\ee
where all derivatives act on the argument of $G$.
Using the relations in appendix~\ref{aanom}, one can write the divergent part as
\be
S^{\rm CT}_{\rm kin,2}=- \frac{1}{\varepsilon} \frac{6 (-1)^{k-1} (k-2)}{(4\pi)^{4k-4} \Gamma(k) \Gamma(3k-2)} g^2
\int \dm^d x \left( \frac{1}{2} \phi (-\Box)^k \phi \right)\,,
\label{kineticCT2}
\ee
which, for $k>2$, leads to the anomalous dimension
\be
\eta =2\gamma= \frac{12 (-1)^{k-1} (k-2)} {\Gamma(k) \Gamma(3k-2)} \frac{g^2}{(4\pi)^{4k-4}} \,, \quad  k>2 \,.
\ee
Substituting the non trivial fixed point value for $g$ given in
Eq.~\eqref{gFP4}, one obtains the universal value
\be
\eta =2\gamma= \frac{3 (-1)^{k-1} (k-2)}{2(4k^2-2k+3)^2} \frac{\Gamma^2(2k)}{\Gamma(k) \Gamma(3k-2)} \varepsilon^2 \,, \quad  k>2 \,.
\label{etauniv}
\ee

The field anomalous dimension is positive for odd values of $k \ge 3$ and goes to zero in the large $k$ limit.
We notice that at two loops there is no divergent counterterm for the $k=2$ case.
In this case the leading part of the ``would be'' counterterm diagram in Eq.~\eqref{kineticCT1} takes, in the $\varepsilon\to 0$ limit,
the finite value
\be
-\frac{3}{4 (4\pi)^2} g^2 \int \dm^d x \left( \frac{1}{2} \phi (-\Box)^2 \phi \right)
\label{finitevalue2loop}
\ee
and there is no contribution at this order to the field anomalous dimension. We shall see that the first contribution comes from a three loop computation.

\subsubsection{Leading order anomalous dimension for \texorpdfstring{$k>2$}{k>2}: alternative derivation with SDE}

One can obtain the scalar field's leading order anomalous dimension
$\gamma$ for the theories in \eqref{theories4} by studying the Schwinger--Dyson equation (SDE) for the two point function~\cite{Rychkov:2015naa,Codello:2017qek}
\be
\langle \phi_x \phi_y \rangle =\frac{\tilde c}{|x-y|^{2\Delta}} \,, \quad \Delta=\delta+\gamma
\ee
obtained from the equation of motion
\be
(-\Box)^k \phi= g \partial_\mu (\partial_\mu \phi \partial_\alpha \phi \partial_\alpha \phi )\,.
\ee
We anticipate that the procedure that we are about to apply does not require the theory to be conformally invariant, but just scale invariant, similarly to RG fixed points,
and thus complements the RG analysis very well.
Our starting point is the relation
\be
(-\Box_x)^{k} (-\Box_y)^{k} \langle \phi_x \phi_y \rangle =\Box_x^{2k} \langle \phi_x \phi_y \rangle = g^2 \partial_{x_\mu} \partial_{y_\nu} \langle (\partial_\mu \phi \partial_\alpha \phi \partial_\alpha \phi )_x
(\partial_\nu \phi \partial_\beta \phi \partial_\beta \phi )_y \rangle\,.
\label{SDE2pt}
\ee
Acting with the derivatives and keeping the leading order contributions, the
left-hand side can be written as
\be
\quad \Box_x^{2k}  \frac{\tilde c}{|x-y|^{2\Delta}} =
\frac{Q_k}{ |x-y|^{2\delta+4k}}\,,
\label{lhsSDE}
\ee
with, defining $F_\Delta=4 \Delta \left( \Delta+1-\frac{d}{2}\right)$,
\be
\quad Q_k=\prod_{i=0}^{2k-1} F_{\Delta+i}=\frac{(-1)^{k-1} 2^{6k-4} k!
(3k-3)!}{(4\pi)^{2k-2} } \gamma \,, \quad k>2\,,
\ee
where, to the leading order, $\tilde c$ has been replaced by the free theory value $c$
and all factors but one, which equates to $\gamma$, are finite for $\varepsilon\to 0$.
Note that, for $k=2$, one obtains $Q_2=-\frac{192}{\pi^2} \gamma $.

At leading order, the right-hand side of Eq.\ \eqref{SDE2pt} is computed considering the correlator for the free theory and performing the Wick contractions.
It reads
\bea
&{}& g^2 \partial_{x_\mu}\partial_{y_\nu}\left[ 2 \langle \partial_\mu \phi_x \partial_\nu \phi_y \rangle (\langle \partial_\alpha \phi_x \partial_\beta \phi_y \rangle)^2
+4 \langle \partial_\mu \phi_x \partial_\beta \phi_y \rangle \langle \partial_\alpha \phi_x \partial_\nu \phi_y \rangle
\langle \partial_\alpha \phi_x \partial_\beta \phi_y \rangle \right]\nonumber \\
\!\!\!\!\!=\, &{}& g^2  \partial_\mu  \partial_\nu \left[ 2 \partial_\mu \partial_\nu G (\partial_\alpha \partial_\beta G)^2 +4 \partial_\mu \partial_\beta G
\,\partial_\beta \partial_\alpha G \,\partial_\alpha \partial_\nu G \right]
= \frac{3 k(k-2) 2^{6k-3} } {(4\pi)^{6k-6} |x-y|^{6k-4} } g^2,
\eea
where $G=G(x-y)$ has been defined in Eq.~\eqref{G} and in the intermediate expression all derivatives on $G(z)$ act on the argument $z$.
The last expression can be derived using
\be
\partial_\mu \partial_\nu G (\partial_\alpha \partial_\beta G)^2=
16 \delta^3 c^3 (k+\delta(3+2\delta)) \frac{2(1+\delta)(x-y)_\mu (x-y)_\nu -(x-y)^2 \delta_{\mu\nu}}{|x-y|^{6\delta+8}}
\ee
and
\be
\partial_\mu \partial_\beta G\,\partial_\beta \partial_\alpha G \,\partial_\alpha \partial_\nu G=
8 \delta^3 c^3 \frac{2(1+\delta)(1+2\delta+4\delta^2) )(x-y)_\mu (x-y)_\nu
-(x-y)^2 \delta_{\mu\nu}}{|x-y|^{6\delta+8}}\,.
\ee
Equating the two sides of the SDE one obtains, for $k>2$, the expression of
the field anomalous dimension $\gamma=\eta/2$
\be
\gamma= \frac{6(-1)^{k-1} (k-2)}{(k-1)! (3k-3)!} \frac{g^2}{(4\pi)^{4k-4}}\,,
\ee
which is in agreement with the perturbative RG computation given above.

We notice that, if one insists in using the SDE for the case $k \to 2$, one gets
the relation with the right-hand side at higher order in $\varepsilon$,
\be
-\frac{192}{\pi^2 |x-y|^8} \gamma =- \frac{9}{ 16 \pi^6 |x-y|^8} g^2
\varepsilon\,,
\ee
which would imply a leading order contribution in $\varepsilon$ and $g$
\be
\lim_{k\to2} \gamma = \frac{3}{4 (4\pi)^4} g^2\varepsilon\,.
\ee
This value is not actually related to the $\varepsilon$ expansion of the anomalous dimension, but rather one can observe that it is related to the finite value of the two loops contribution after multiplying and dividing it by $\varepsilon$ and treating it as a coefficient (proportional to $\varepsilon$) times a $1/\varepsilon$ factor, as if it was UV divergent.

%

\subsubsection{Leading order anomalous dimension for \texorpdfstring{$k=2$}{k=2}: perturbative RG approach}

For $k=2$ the scalar field has zero canonical dimension and does not acquire an anomalous dimension at two loops,
so that we need to examine the contributions at three loops.
The counterterms of order $g^3$ to the kinetic term in the Euclidean action, which define the field strength, are represented diagrammatically in Fig.~\ref{CTkinLO4}-(b).

The first class of contributions can be written as
\be
S^{\rm CT}_{\rm kin,3,1}=- 2 \frac{1}{2} \frac{1}{2^2} \int \prod_{i=1,10} \dm^d x_i \,S^{(3)} G G \,S^{(4)} GG \,S^{(3)} G\,,
\ee
where $S^{(4)}$ can be trivially obtained by a functional derivative on $S^{(3)}$ given in Eq.~\eqref{S3} and $G$ is as usual the free propagator.
Each vertex has three independent terms and the $27$ terms in the product
are not all independent, but instead can be grouped in $7$ classes:
\be
S^{\rm CT}_{\rm kin,3,1}=\left(\frac{g}{4}\right)^3 \frac{8^3}{4}\left(2 D_1+8D_2+4D_3+4D_4+D_5+4D_6+4D_7\right)\,.
\ee
Looking at the UV divergent part with the $1/\varepsilon$ pole, we find
\be
\left(D_1,D_2,D_3,D_4,D_5,D_6,D_7\right)_{\rm div}=\left(
\frac{5}{36}, \frac{1}{16}, \frac{7}{144}, \frac{7}{144}, \frac{5}{9},  \frac{2}{9}, \frac{11}{144}
\right)\! \frac{1}{(4\pi)^6\varepsilon}\int \dm^d x \frac{1}{2} \phi (-\Box)^2 \phi\,, \nonumber
\ee
and therefore
\be
S^{\rm CT}_{\rm kin,3,1}=\frac{35}{6}\frac{g^3}{(4\pi)^6}\frac{1}{\varepsilon}\int \dm^d x \frac{1}{2} \phi (-\Box)^2 \phi\,.
\ee

The second contribution in Fig.~\ref{CTkinLO4}-(b) is obtained from the two loop diagrams where one vertex is replaced by its one loop counterterm. In practice, it can be obtained combining the results in Eqs.~\eqref{finitevalue2loop} and~\eqref{vertexCT} for $k=2$ and reads
\be
S^{\rm CT}_{\rm kin,3,2}=-\frac{15}{2}\frac{g^3}{(4\pi)^6}\frac{1}{\varepsilon}\int \dm^d x \frac{1}{2} \phi (-\Box)^2 \phi\,.
\ee
Summing these two contributions together, we have that at order $g^3$ the leading order kinetic counterterm becomes
\be
S^{\rm CT}_{\rm kin,3}=-\frac{5}{3}\frac{g^3}{(4\pi)^6}\frac{1}{\varepsilon}\int \dm^d x \frac{1}{2} \phi (-\Box)^2 \phi \,,
\ee
which gives the wavefunction renormalization $Z=1-\frac{1}{\varepsilon} \frac{5}{3}\frac{g^3}{(4\pi)^6}$
and leads to the anomalous dimension at leading order
\be
\eta=5 \frac{g^3}{(4\pi)^6}=\frac{1}{25}\varepsilon^3\,,
\ee
where we used the definition $\eta=\beta_g \frac{d \log Z}{d g}$.

\section{Cubic models: \texorpdfstring{$\boldsymbol{\varepsilon}$}{epsilon}-expansion analysis} \label{Cubic_theories}

We consider now the simplest family of critical shift invariant theories with purely cubic interactions, which, after performing the Wick's rotation like in the previous section, is written in terms of the Euclidean action $S=S_E$
\be
S=\int {\rm d}^{d}x \left[ \frac{1}{2}\phi \left(-\Box\right)^k \phi +\frac{g}{2} (\partial \phi \cdot \partial \phi) \Box \phi \right]\,,
\label{theories3}
\ee
below the upper critical dimension $d_c=2(3k-4)$ at $d=d_c-\varepsilon$.

The propagator of the massless free theory (for $g=0$) in the coordinate
representation has already been given in the previous section in Eq.~\eqref{G},
but now $\delta=2(k-2)-\frac{\varepsilon}{2}$. The equation of motion
from which one builds the Schwinger--Dyson equations is given by
\be
(-\Box)^k \phi= - \frac{\delta S_{\rm int}}{\delta \phi(x)} = g \left[ (\Box \phi) ^2 -(\partial_\mu \partial_\alpha \phi)^2 \right]\,.
\label{cubicEOM}
\ee
The cubic interaction operator $(\partial \phi \cdot \partial \phi) \Box \phi$ is
relevant from the point of view of the free theory since
$\Delta_{(\partial \phi \cdot \partial
\phi) \Box \phi}=d-\frac{\varepsilon}{2}<d$.

\subsection{Leading order anomalous dimension}

To begin with, we compute the scalar field anomalous dimension $\gamma$ at the leading
order for the theories in \eqref{theories3}.
One can achieve this most simply by studying the SDE for the two point function
\be
\langle \phi_x \phi_y \rangle =\frac{\tilde c}{|x-y|^{2\Delta}} \,, \quad \Delta=\delta+\gamma\,.
\ee
Applying the $k$-th power of the Laplacian in both $x$ and $y$, on can use the equation of motion~\eqref{cubicEOM} and Wick contractions to obtain the relation
\bea
(-\Box_x)^{k} (-\Box_y)^{k} \langle \phi_x \phi_y \rangle &=&\Box_x^{2k} \langle \phi_x \phi_y \rangle =
g^2 \langle  \left[ (\Box \phi) ^2 -(\partial_\mu \partial_\alpha \phi)^2 \right]_x
 \left[ (\Box \phi) ^2 -(\partial_\nu \partial_\beta \phi)^2 \right]_y \rangle \nonumber \\
 &=& 2g^2\left[
 (\Box^2 G)^2-2 (\de_\mu \de_\alpha \Box G)^2 +(\de_\mu \de_\nu \de_\alpha \de_\beta G)^2
 \right]\,.
\label{cubicSDE2pt}
\eea

Evaluating the left hand side at leading order, likewise the even case, one finds
\be
 \Box_x^{2k}  \frac{\tilde c}{|x-y|^{2\Delta}} =
\frac{Q'_k}{ |x-y|^{2\delta+4k}}\,,
\label{lhsSDEcubic}
\ee
with the constant $Q'_k$ defined as
\be
 Q'_k=\prod_{i=0}^{2k-1} F_{\Delta+i}=\frac{(-1)^{k-1} 4^{4k-4} k!
(4k-5)!}{(4\pi)^{3k-4} } \gamma \,, \quad k>2\,,
\ee
where, to the leading order, $\tilde c$ has been replaced by the free theory value $c$ and all factors except for one, that is equal to $\gamma$, are finite when $\varepsilon\to 0$.

The right-hand side of the SDE equation can be computed from the explicit expression of the green function $G$ in the coordinate representation
and reads
\be
2^9 3 g^2 c^2 k (k-1) (k-2)^2 (2k-3)^3 (6k-7) \frac{1}{
|x-y|^{2\delta+4k}}\,,
\ee
so that one also obtains the expression for the anomalous dimension
\be
\gamma=\frac{3 (-1)^{k-1} }{2 (4\pi)^{3k-4} }
\frac{(2k-3) (6k-7) \Gamma^2(2k-2)}{\Gamma(k-1) \Gamma^2(k) \Gamma(4k-4)} g^2
= \gamma^{(1)} g^2\,,
\label{gamma_cubic}
\ee
which can be manipulated to the slightly more compact form
\be
 \gamma^{(1)}= \frac{3 (-1)^{k-1} }{4^k (4\pi)^{3k-4} }  \frac{(2k-3)^2 (6k-7) \Gamma(k-\frac{3}{2})}{ \Gamma^2(k) \Gamma(2k-\frac{3}{2})}\,.
 \label{gamma1}
\ee

One can derive the same anomalous dimension with perturbative RG methods computing the counterterm to the field strength,
starting from the general expression already presented in
Eq.~\eqref{generic_bubble}~\footnote{%
Indeed plugging into this expression the two different second derivatives of the action for the quartic and cubic families, one obtains the corrections to the 4pt- and 2pt- functions, respectively.
}
and plugging the cubic interaction encoded in
\be
\frac{\delta^{2} S_{\rm int}}{\delta \phi(x_1) \delta \phi(x_2)}= g \int \dm^d x_0 \left[
\Box \phi \de_\mu \delta_{x_{01} }   \de_\mu \delta_{x_{02} } +
\de_\mu \phi  \de_\mu \delta_{x_{01} } \Box \delta_{x_{02} } +
\de_\mu \phi  \de_\mu \delta_{x_{02} } \Box \delta_{x_{01} }
\right] \,.
\label{S2cubic}
\ee
After calculations that are somewhat more involved than the previous section,
one finds
divergent contributions from four types of diagrams.
Summing them together one can construct the perturbative field strength $Z$ and therefore compute
$\eta=2\gamma=-\frac{\varepsilon}{2} g \frac{d}{d g} \log{Z}$, finding complete agreement with Eq.~\eqref{gamma_cubic}.

\subsection{One loop perturbative \texorpdfstring{$\beta$}{beta} function}

\begin{figure}
      \scalebox{2}{
       \begin{tikzpicture}[baseline=-.1cm]
     \draw (0,0) circle (.5cm);
     \filldraw [gray!100] (0,.5) circle (2pt);
     \draw (0,.5) circle (2pt);
     \filldraw [gray!100] (.469,-.171) circle (2pt);
     \draw (.469,-.171) circle (2pt);
     \filldraw [gray!100] (-.469,-.171) circle (2pt);
     \draw (-.469,-.171) circle (2pt);
     \end{tikzpicture}
     }

        \caption{Diagram in coordinate representation for the one loop corrections to the cubic vertex.}
        \label{CTint1loop3}
 \end{figure}
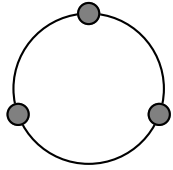

The $\overline{\mathrm{MS}}$ RG flow, which is needed to determine the fixed point value of the critical couplig $g$, is given by
\be
\mu \frac{d g}{d\mu} = \beta_3(g)= \left( -\frac{\varepsilon}{2}+3\gamma
\right) g +a_1 g^3=-\frac{\varepsilon}{2} g+ (3\gamma^{(1)} +a_1 ) g^3\,.
\label{beta_cubic}
\ee
The coefficient $a_1$ is determined by computing the one loop correction to the cubic vertex,
corresponding to the diagram shown in Fig.~\ref{CTint1loop3}.

More precisely, let us write the one loop counter-term of the cubic interaction,
picking up as usual the divergent part, in the coordinate representation
\be
S^{\rm CT}_{\rm int,3}=- \int \prod_{i=1}^6 \dm^d x_i \, \frac{1}{3!}  \left[  \frac{\delta^{2} S_{\rm int}}{\delta \phi(x_6) \delta \phi(x_1)}
G_{x_{12}}
\frac{\delta^{2} S_{\rm int}}{\delta \phi(x_2) \delta \phi(x_3)}
G_{x_{34}}
\frac{\delta^{2} S_{\rm int}}{\delta \phi(x_4) \delta \phi(x_5)}
G_{x_{56}}  \right]_{\rm div}\,,
\label{cubicintCT}
\ee
where $x_{ij}=x_i-x_j$. The Hessian $\frac{\delta^{2} S_{\rm int}}{\delta \phi(x_1) \delta \phi(x_2)}$ was already given in Eq.~\eqref{S2cubic}.

In the complete expression of Eq.~\eqref{cubicintCT} one has a total of $27$ contributions which can be arranged in $6$ distinct groups as
\be
S^{\rm CT}_{\rm int,3}= \frac{g^3}{3!} \left(\sum_{i=1}^9 E_{1i}+\sum_{i=1}^3E_{2i}+\sum_{i=1}^6E_{3i}
+\sum_{i=1}^2E_{4i}+\sum_{i=1}^6E_{5i}+E_6\right)_{\rm div}\,.
\label{cubicsumCT}
\ee
First of all, one can notice that the last two terms, in the groups $E_5$ and $E_6$, are independently finite, i.e.\ they do not affect the coefficient $a_1$ of the beta function in Eq.~\eqref{beta_cubic}, therefore we do not need to compute them.
We find (see Appendix~\ref{betacubic}) that in the group $E_1$ all contributions have the same absolute values but $6$ are negative and $3$ are positive,
in the group $E_2$ all contributions are the same,
in the group $E_3$ there are $6$ equal contributions,
and, finally, in the group $E_4$ one has two contributions which are equal.
Concretely, we have
\bea
 \left(\sum_{i=1}^9 E_{1i}, \sum_{i=1}^3E_{2i}, \sum_{i=1}^6E_{3i},
\sum_{i=1}^2E_{4i}, \sum_{i=1}^6E_{5i}, E_6\right)_{\rm div} &=&\nonumber  \\
&{}& \hspace{-7cm}  \left(6,-6,-6, 6 ,0,0\right)
\frac{1}{(4\pi)^{3k-4} \Gamma(3k\!-\!3)}\frac{1}{\varepsilon}  \int \dm^d x
\frac{1}{2} (\partial \phi \cdot \partial \phi) \Box \phi\,.
\eea
Surprisingly, after summing over all contributions,
one discovers that there is a complete cancellation:
\be
S^{\rm CT}_{\rm int,3}= \frac{a_1}{\varepsilon} g^3 \int \dm^d x \frac{1}{2} (\partial \phi \cdot \partial \phi) \Box \phi   \,, \quad
a_1=0
\,.
\ee

The source of this cancellation in this family of cubic theories in unclear,
but it is probably related to the shift symmetry, which constrains the interaction to be given by just the operator considered above, in combination with the cubic nature of the interactions. This fact leads to a rather peculiar form of the beta function, which determines quantum scale invariance, and is similar to what happens in some SUSY theories, which enjoy non renormalization conditions.
In practice, the beta function depends only on a linear scaling term plus the anomalous dimension contribution of cubic order in the coupling.

In particular, at the fixed point, we find the criticality conditions
\be
g=\pm  \sqrt{ \frac{\varepsilon} {2 (3\gamma^{(1)} +a_1 )}  }=\pm  \sqrt{ \frac{\varepsilon} {6 \gamma^{(1)} }  }\,,
\label{cubicgsol}
\ee
where $\gamma^{(1)}$ is given in \eqref{gamma1}. As a consequence, $g$ is either real or purely
imaginary, for odd or even values of $k$ respectively. The two solutions,
either with positive or negative sign in \eqref{cubicgsol},
are  physically equivalent (because we can map one to the other by redefining $\phi\to-\phi$)
and they correspond to an attractive fixed point since
$\Delta_{(\partial\phi\cdot\partial\phi)\square\phi}=d+\partial_g\beta_3(g)
|_{g~{\rm of}~\eqref{cubicgsol}}=d+\varepsilon>d$.
For this family of critical theories then the leading order universal anomalous dimension of
the field $\phi$ is $k$-independent and positive:
\be
\gamma=\frac{\varepsilon}{6} \, , \qquad k\ge 3\,.
\ee

%

\section{Energy momentum tensor and conformal invariance}\label{EMT_CFT}

One may ask if the kind of quantum theories described above, and in
particular the two families investigated within perturbation theory in
Sections~\ref{Quartic_theories} and~\ref{Cubic_theories}, are only scale or
also conformally invariant at criticality (as a fixed point of the RG flow).

Let us start by summarizing some well-known properties of QFTs that are
relevant for our discussion. Denoting the action of the infinitesimal
global scale variation by $\delta_\omega$ on the line element (changing
coordinates or equivalently the metric) and on the fields, one generically
obtains a change in the quantum effective action given by the trace
anomaly,
\be
\delta_\omega \Gamma=-\omega \int_x \langle
\eta_{\mu\nu}T^{\mu\nu}(x)
\rangle,
\ee
with $\eta_{\mu\nu}$ the mostly-plus Minkowski metric, $T^{\mu\nu}$ the
energy momentum tensor (EMT), and the quantum contribution to the trace of
the EMT is
\be
\eta_{\mu\nu}T^{\mu\nu}|_{\rm quant.}
=\sum_i \tilde{\beta}_i(\lambda_i) \frac{\partial {\mathscr
L}}{\partial \lambda_i} \,.
\ee
The $\tilde \beta_i$ are the beta functions of the couplings multiplying
the various interaction operators.\footnote{We use the notation
$\tilde{\beta}_i$ because, differently from the $\beta_i$ obtained in
previous section, these beta functions do not contain the classical term
proportional to $\varepsilon$. The relation between $\tilde{\beta}_i$ and
$\beta_i$ should become more transparent in the explicit computations of
the expectation value of the trace of the EMT given below.} They can be
obtained with several methods, e.g.\ the ones we saw in the two previous
sections for the two families of theories investigated there.

A global scale variation leads generically to an additional term given by a
total divergence $\partial_\mu K^\mu$. The conserved Noether dilatation current $D^\mu$ is given by
\be
D^\mu=x_\nu T^{\mu\nu} +K^\mu  \quad  \Longrightarrow \quad
\eta_{\mu\nu}T^{\mu\nu} +\partial_\mu K^\mu=0 \,.
\ee
The vector $K^\mu$ is known as the virial current and can be defined in any
scale invariant theory.  If the theory is also conformal, additional
requirements arise from the invariance under special conformal
transformations so that the virial current takes the form of the divergence
of a tensor,
\be
K^\mu = \partial_\nu L^{\mu\nu}\,.
\label{confcond}
\ee
%


In general the EMT obtained via the standard Noether procedure is not
symmetric, but it can always be improved following the Belinfante
procedure. However, the easiest way to obtain improved forms of symmetric
and conserved EMT is to consider the QFT of interest in a curved
background, where we add appropriate local interaction terms which are zero
in a flat background.  They typically contribute to the trace of the EMT as
a double divergence and therefore affect the form of the tensor
$L^{\mu\nu}$ introduced above, as can be seen in the couplings to the Ricci
tensor and the scalar curvature.  However, more general terms can appear
and be of interest. For example it has been shown in~\cite{Osborn:2016bev}
that for theories with a cubic Laplacian kinetic term, in order to have an
EMT which is a conformal primary, one needs to add a term in the curved
space action involving the Bach tensor, which does not affect the trace of
the EMT.

Let us now analyze the two families of QFTs with quartic and cubic
interaction to see if they are scale or conformally invariant at the fixed
point which was found in perturbation theory. We will compute the EMT
explicitly. We will not do this with the curved-space method, but rather
follow the method described in~\cite{OsbornLectures}, which can be used
whenever a CFT is derived from an action. Here we will present the relevant
results of~\cite{OsbornLectures} and the reader is invited to
peruse~\cite{OsbornLectures} where the method is described in detail.

Under a conformal transformation $x\to x'$, infinitesimal line elements
remain invariant up to a local scale factor, i.e.\ ${\rm
d}x'^{\,2}=\Omega^2(x){\rm d}x^2$, ${\rm d}x^2=\eta_{\mu\nu}{\rm d}x^\mu
{\rm d}x^\nu$. A scalar field $\phi=\phi(x)$ transforms as
\begin{equation}
  \phi(x)\to\phi'(x')=\Omega(x)^{-\Delta_\phi}\phi(x)\,,
  \label{conftrans}
\end{equation}
where $\Delta_\phi$ is the scaling dimension of $\phi$. For infinitesimal
transformations, $\delta x^\mu=x^{\prime\,\mu}-x^\mu=v^\mu(x)$,
$\Omega(x)=1+\sigma(x)$, we find
\begin{equation}
  \delta_{v,\sigma}\phi=-v^\mu\partial_\mu\phi-\sigma\,\Delta_\phi\,\phi\,.
  \label{infconftrans}
\end{equation}
Then, for an action $S[\phi]$ we have
\begin{equation}
  \begin{aligned}
    \delta_{v,\sigma}S[\phi]=\int {\rm d}^d x\,&\big[(\partial_\mu
    v_\nu-\sigma\,\eta_{\mu\nu})T_c^{\mu\nu}-(\partial_\mu\sigma\,
    \eta_{\nu\rho}-\partial_\nu\sigma\,\eta_{\mu\rho})X^{\rho\mu\nu}\\
    &+(\partial_\nu\partial_\rho v_\mu +\partial_\mu\sigma\,\eta_{\nu\rho}
    - \partial_\nu\sigma\,\eta_{\mu\rho}-\partial_\rho\sigma\,\eta_{\mu\nu})
  Y^{\mu\rho\nu}+\partial_\mu\partial_\nu \sigma\, Z^{\mu\nu}\big]\,,
  \end{aligned}
  \label{varaction}
\end{equation}
where $X^{\rho\mu\nu}=-X^{\rho\nu\mu}$, $Y^{\mu\rho\nu}=Y^{\mu\nu\rho}$ and
$Z^{\mu\nu}=Z^{\nu\mu}$. The right-hand side of \eqref{varaction} vanishes
on-shell for infinitesimal conformal transformations if the action is
conformally invariant. In this context, the symmetric and conserved EMT of
the CFT is given by\footnote{A term $\partial_\rho X^{\rho\mu\nu}$ is absent compared to the general form given in~\cite{OsbornLectures} because the field $\phi$ is a Lorentz scalar.}
\be
T^{\mu\nu}=T^{\mu\nu}_c+\partial_\rho(
-X^{\mu\rho\nu}
-X^{\nu\rho\mu} +Y^{\rho\mu\nu} -Y^{\mu\nu\rho} -Y^{\nu\mu\rho})
+\mathcal{D}^{\mu\nu\rho\sigma}Z_{\rho\sigma}\,,
\label{improvedT}
\ee
 with
\begin{equation}
  \begin{aligned}
    \mathcal{D}^{\mu\nu\rho\sigma}&=\frac{1}{d-2}
    \big(\eta^{\mu(\rho}\partial^{\sigma)}\partial^\nu
    +\eta^{\nu(\rho}\partial^{\sigma)}\partial^\mu
    -\eta^{\mu(\rho}\eta^{\sigma)\nu}\square
    -\eta^{\mu\nu}\partial^\rho\partial^\sigma\big)\\
    &\quad+\frac{1}{(d-2)(d-1)}(\partial^\mu\partial^\nu
    -\eta^{\mu\nu}\square)\eta^{\rho\sigma}\,,
  \end{aligned}
\end{equation}
which satisfies $\partial_\mu\mathcal{D}^{\mu\nu\rho\sigma}=0$ and
$\eta_{\mu\nu}\mathcal{D}^{\mu\nu\rho\sigma}=-\partial^\rho\partial^\sigma$.
$T^{\mu\nu}$ given in \eqref{improvedT} is conserved, symmetric and
traceless subject to the equations of motion.  We observe that there is
still the freedom to add to this improved expression of the EMT a further
term conserved and of zero trace, analogous to the last one in
\eqref{improvedT}, but with $\mathcal{D}^{\mu\nu\rho\sigma}$ replaced by a
suitable $\mathcal{D}_B^{\mu\nu\rho\sigma}$  such that
$\partial_\mu\mathcal{D}_B^{\mu\nu\rho\sigma}=0$ and
$\eta_{\mu\nu}\mathcal{D}_B^{\mu\nu\rho\sigma}=0$. This has been shown to
be needed in higher derivative free theories to make the EMT  operator a
conformal primary~\cite{Osborn:2016bev}.

Our task, now, is to compute $T^{\mu\nu}$ by explicitly evaluating the
variation $\delta_{v,\sigma}S[\phi]$ for the actions in
Sections~\ref{Quartic_theories} and~\ref{Cubic_theories}. Before that, let
us briefly see how this method works in the standard $\phi^4$ theory with action
\eqn{S_{\phi^4}[\phi]=-\int {\rm d}^dx
\left(\frac12\partial^\mu\phi\lsp\partial_\mu\phi
+\frac{\lambda}{4!}\phi^4\right)\,.}[phifourth]

In $d=4-\varepsilon$ the classical scaling dimension of $\phi$ is
$\Delta_\phi=\tfrac12(d-2)=1 - \frac12\varepsilon$. Of course the theory
\phifourth is not classically conformally invariant and so we expect
\eqn{\delta_\sigma S_{\phi^4}[\phi]
=\varepsilon\lsp\frac{\lambda}{4!}
\int{\rm d}^d x\,\sigma\lsp\phi^4}[]
due to the classical scaling of the coupling $\lambda$. The contributions
of the kinetic term to the EMT have already been computed
in~\cite{OsbornLectures}. Including the contribution arising from the
interaction term, we find\foot{Here and below tensors that are not listed are zero.}
\eqn{T^{\mu\nu}_c=\partial^\mu\phi\lsp\partial^\nu\phi
-\frac12\eta^{\mu\nu}\lsp\partial^\rho\phi\lsp\partial_\rho\phi
-\frac{\lambda}{4!}\eta^{\mu\nu}\phi^4\,,\qquad
Z^{\mu\nu}=-\tfrac14(d-2)\eta^{\mu\nu}\lsp\phi^2\,,}[]
and so
\eqn{T^{\mu\nu}=\partial^\mu\phi\lsp\partial^\nu\phi
-\frac12\eta^{\mu\nu}\lsp\partial^\rho\phi\lsp\partial_\rho\phi
-\frac{\lambda}{4!}\eta^{\mu\nu}\phi^4
-\frac{d-2}{4(d-1)}(\partial^\mu\partial^\nu-\eta^{\mu\nu}\square)
\phi^2\,.}[]
From this expression and upon using the equation of motion,
$\square\phi=\frac{\lambda}{3!}\phi^3$, we find $\eta_{\mu\nu}
T^{\mu\nu}=-\varepsilon\frac{\lambda}{4!}\phi^4$ as expected. In the quantum
theory there is an additional beta function contribution (trace anomaly)
giving the one-loop quantum EMT trace as the well-known expression
$
\eta_{\mu\nu}T^{\mu\nu}|_{\rm one~loop}
=\left(-\varepsilon \lambda+3\frac{\lambda^2}{(4\pi)^2} \right)\frac{1}{4!}\phi^4$. This allows
for a zero EMT trace, and thus a conformal theory, at the associated fixed
point.

\subsection{Shift symmetric theories with a quartic interaction}

Considering the one with quartic interaction with action in flat space
\be
S_4=-\int {\mathscr L}_4 =- \int {\rm d}^{d}x \left[ \frac{1}{2}\phi \left(-\Box\right)^k \phi +\frac{g}{4} (\partial \phi \cdot \partial \phi)^2\right]\,,
\label{quartic_theories}
\ee
we need to compute the trace of the energy momentum tensor. We will focus
on the contributions to the EMT arising from the interaction term. We are
not aware of a general result (for all $k$) for the contributions to the
EMT arising from the kinetic term, but the contribution of the kinetic term
to the trace of the EMT is conjectured to be given by
\eqn{\eta_{\mu\nu}T^{\mu\nu}_{\text{kin}}=-\Delta_\phi\,
\phi(-\square)^k\phi\,,\qquad\Delta_\phi=\tfrac12(d-2k)\,,}
a result based on the $k=2,3$ cases computed explicitly
in~\cite{Osborn:2016bev} and consistent with the $k=4$ case computed on-shell in
\cite{Guerrieri:2016whh}.\footnote{
For completeness, we also conjecture the general formula for the divergence
$\partial_\mu T^{\mu\nu}_{\text{kin}}=-\partial^\nu \phi (-\square)^k\phi $,
which is also verified for $k=2,3$ and consistent with $k=4$.
Both $\eta_{\mu\nu}T^{\mu\nu}_{\text{kin}}$ and $\partial_\mu T^{\mu\nu}_{\text{kin}}$
are proportional to the equation of motion of the free theory and thus become zero on shell in that case.
}

For the interaction term we find
\eqna{T_{c,\text{int}}^{\mu\nu}&=
g\lsp\partial^\mu\phi\lsp\partial^\nu\phi\lsp\lsp
\partial\phi\cdot\partial\phi
-\tfrac14\lsp g\lsp\eta^{\mu\nu}(\partial\phi\cdot\partial\phi)^2\,,\\
X^{\rho\mu\nu}_\text{int}&=-\frac{1}{2(d-1)}\lsp
g\lsp\Delta_\phi\phi^2\eta^{\rho[\mu}
\partial^{\nu]}(\partial\phi\cdot\partial\phi)\,,\\
Z^{\mu\nu}_\text{int}&=-\tfrac12 \lsp g\lsp\Delta_\phi\eta^{\mu\nu}
\phi^2\lsp\partial\phi\cdot\partial\phi\,.}[]
Using these results in the general form \eqref{improvedT} and applying
the equation of motion $(-\square)^{k}\phi=g\lsp\partial^\mu(\phi\lsp\partial_\mu\phi\lsp\lsp
\partial\phi\cdot\partial\phi)$, we find
in $d=4(k-1)-\varepsilon$ dimensions
\eqn{\eta_{\mu\nu}T^{\mu\nu}=\eta_{\mu\nu}(T^{\mu\nu}_{\text{kin}}
+T^{\mu\nu}_{\text{int}})=-\varepsilon\frac{g}{4}\lsp
(\partial\phi\cdot\partial\phi)^2\,.}[]

The contribution induced by the trace anomaly, according to Eq.~\eqref{vertexCT}, is given by
\be
\tilde{\beta}_4 \frac{\partial {\mathscr L}_4}{\partial g} = 2 \frac{4k^2-2k+3}{(4\pi)^{2k-2}\Gamma(2k)} g^2 \frac{1}{4} (\partial\phi \cdot \partial\phi )^2\,
\label{traceanomaly4}
\ee
which at quantum level, at one loop, leads to the total trace of the EMT
\be
\eta_{\mu\nu}T^{\mu\nu}|_{\rm one~loop}=
\left( -\varepsilon g+2 \frac{4k^2-2k+3}{(4\pi)^{2k-2}\Gamma(2k)} g^2\right) \frac{1}{4} (\partial\phi \cdot \partial\phi )^2\,,
\ee
which vanishes at the fixed point. From this one loop analysis one expects that the quantum critical theory is conformal and not only scale invariant.

\subsection{Shift symmetric theories with a cubic interaction}

We consider now the $k$-family of shift symmetric theories with cubic interactions with the action
\be
S_3=-\int {\mathscr L}_3=-\int {\rm d}^{d}x \left[ \frac{1}{2}\phi \left(-\Box\right)^k \phi +\frac{g}{2} (\partial \phi \cdot \partial \phi) \Box \phi \right]\,,
\label{cubic_theories}
\ee
with $d=d_c-\varepsilon$ and $d_c=2(3k-4)$,
for which we have shown previously that the contribution to the trace anomaly is given by
\be
\tilde{\beta}_3 \frac{\partial {\mathscr L}_3}{\partial g} = 3 \gamma^{(1)} g^3 \frac{1}{2} \int \dm^d x(\partial\phi \cdot \partial\phi ) \Box \phi\,.
\label{traceanomaly3}
\ee

Focusing again on the interaction term, we find
\eqna{T_{c,\text{int}}^{\mu\nu}&=g\lsp\partial^\mu\phi\lsp
\partial^\nu\phi\lsp\lsp\square\phi
-\tfrac12\lsp g\lsp\eta^{\mu\nu}\partial\phi\cdot\partial\phi\,\square\phi
+g\lsp\partial\phi\cdot\partial\phi\,\partial^\mu\partial^\nu\phi\,,\\
X^{\rho\mu\nu}_{\text{int}}&=-\frac{1}{2(d-1)}\lsp g\big[\Delta_\phi\lsp
\phi^2\eta^{\rho[\mu}\lsp\partial^{\nu]}\square\phi-
(2\lsp\Delta_\phi-d-2)\partial\phi\cdot\partial\phi\,\eta^{\rho[\mu}
\partial^{\nu]}\phi\big]\,,\\
Y^{\mu\rho\nu}_{\text{int}}&=\tfrac12\lsp g\lsp \eta^{\rho\nu}
\partial\phi\cdot\partial\phi\,\partial^\mu\phi\,,\\
Z^{\mu\nu}_{\text{int}}&=-\tfrac12\lsp g\lsp\Delta_\phi\eta^{\mu\nu}
(\phi^2\lsp\square\phi-\phi\lsp\partial\phi\cdot\partial\phi)\,.}[]
With use of the general form of the improved EMT \eqref{improvedT}
and the equation of motion $(-\square)^k\phi=g[(\square\phi)^2 -
\partial_\mu\partial_\nu\phi\lsp\partial^\mu\partial^\nu\phi]$, in
$d=2(3k-4)-\varepsilon$ dimensions we obtain
\eqn{\eta_{\mu\nu}T^{\mu\nu}=\eta_{\mu\nu}(T^{\mu\nu}_{\text{kin}}
+T^{\mu\nu}_{\text{int}})=-\frac{\varepsilon}{2}\frac{g}{2}\lsp
\partial\phi\cdot\partial\phi\,\square\phi\,.}[]
Taking into account the trace anomaly,
one has, at quantum level and at one loop, the total trace of the EMT
\be
\eta_{\mu\nu}T^{\mu\nu}|_{\rm one~loop}=
\left( -\frac{\varepsilon}{2} g+\gamma^{(1)}  g^3 \right) \frac{1}{2}\lsp \partial\phi\cdot\partial\phi\,\square\phi \,,
\ee
with $\gamma^{(1)}$ given in eq.~\eqref{gamma1}. Therefore the quantum theory of the model with cubic interaction is also conformal at the fixed point in the one loop approximation.
\section{Conclusions}

In this work we have considered higher derivative single scalar field theories
with shift symmetric derivative interactions. First we have given a rather general
classification of the possible critical theories below their corresponding upper critical dimensions.

For the two simplest families, representative of the two cases with a
single even and a single odd classically marginal interaction, we
have investigated the critical theories at leading order in perturbation theory
in the $\varepsilon$ expansion, computing the respective beta functions,
the field's anomalous dimensions
and, for the even case, also the scaling dimension of the most relevant
deforming operator that respects the shift symmetry. The field anomalous
dimension has been obtained with both RG and SDE methods using perturbation theory.
An interesting outcome of our computation is that, for the family of cubic
theories, investigated in Section~\ref{Cubic_theories},
the interaction vertex
does not renormalize. If this persists to higher orders and is not an accidental
feature, it should be an outcome of the interplay of the shift symmetry with the odd
nature of the interation that should be understood better.

The perturbative RG analyses of Sections~\ref{Quartic_theories}
and~\ref{Cubic_theories} have shown that the two shift symmetric families of
theories, even and odd, at the fixed point of the RG flow are scale
invariant. Therefore, in the last section, we have adopted a generalized
Noether approach, described in~\cite{OsbornLectures}, with which it is possible to
construct improved energy-momentum tensors with a zero trace at the fixed
points of the RG flow of the models.
The implication is that these critical theories are actually
conformally invariant at least at one loop.
Interestingly, in these theories scale invariance
implies conformal invariance even though unitarity is
absent.\footnote{%
Scale without conformal invariance requires the presence
of a spin-one, dimension $d-1$, operator that is not conserved. It is not
clear what would prevent such an operator from obtaining an anomalous dimension
in the presence of interactions. Therefore, even without unitarity, finding
an interacting theory with such an operator appears to be a highly
nontrivial task.
}
However, we have not presented the perturbative construction of
the renormalized energy momentum tensor,
which in general needs further improvements to
become a conformal primary. This would certainly deserve a detailed
treatment which goes beyond the scope of this work.

The extent in which these critical theories actually
exist in physical lower dimensions (below their upper critical dimensions)
is an open
question, that reasonably requires non perturbative investigations.
We expect some of these theories to exist in $d=4$ as CFTs,
likely keeping their non unitary nature.
Therefore, it would be a natural task, but certainly a non trivial one, to
carry out an analysis based on the conformal bootstrap in this direction.

Shift symmetric theories can be useful when constructing models
of extended objects, such as strings and branes,
both with intrinsic and extrinsic geometrical properties.
In these cases, the scalar field $\phi$ is replaced by a space(time) embedding
function $X^\mu$ and the coordinates $x^\mu$
are charts on the extended objects themselves.
An incomplete list of extended objects includes Polyakov's string,
which has recently been generalized into a higher derivative model
that shares some similarities with those discussed in this paper
\cite{Romoli:2021hre},
but also extrinsic membrane theories such as Helfrich's and tethered membrane
that have a long history \cite{David:1992vv,David:1993rr},
but are still a fertile area of research \cite{Coquand:2020tgb,Metayer:2021kxm}.

We note that there are other kind of QFTs that can be constructed and
studied in the same spirit as the ones presented here.
A first straightforward example
would be a possible generalization obtained by introducing fermionic
fields. However, even concentrating on purely scalar theories, also shift invariant
multiscalar field theories can be of interest, both without assuming any
global symmetry, so that one could classify the critical theories and the
symmetries allowed at the RG fixed point, or considering some specific
symmetry group. Recently, several investigations have been devoted to the
problem of finding novel critical theories and the emergence of global
symmetries at criticality~\cite{Osborn:2017ucf,Codello:2019isr,
Codello:2020lta, Osborn:2020cnf} and the relations to CFT in a perturbative
context~\cite{Codello:2018nbe}.

As an example of a possible extension,
we mention the generalization to the multifield case
$\{\phi^i\}, i=1,\ldots,N,$ of the family of theories investigated in
Section~\ref{Quartic_theories}, given by generic marginal interactions
encoded in
\be
V=\frac{1}{4} \sum_{p=1}^3 \lambda^{(p)}_{ijkl} (\partial \phi^i \cdot  \partial \phi^{j_p}) (\partial \phi^{k_p} \cdot  \partial \phi^{l_p}) \,,
\ee
where $p$ gives the three independent cyclic permutations $(j_p,k_p,l_p)$
for the indices $(j,k,l)$. There is a similar extension for the family with
cubic interactions.  A special case would be given by the $O(N)$ symmetric
theory with multiplet $\vec{\phi}$, where the potential reduces to
\be V=\lambda_1  (\partial \vec \phi \cdot  \partial \vec \phi)^2 +
\lambda_2(\partial_\mu \vec \phi \cdot \partial_\nu \vec \phi)
(\partial^\mu \vec \phi \cdot  \partial^\nu \vec \phi)\,,
\label{ONpot}
\ee
for which one could study analytically the large $N$ behavior. Theories
with higher derivative kinetic terms but non shift symmetric interactions
have been considered in~\cite{Gracey:2015xmw, Gracey:2017erc}.  Another
possible extension of our analysis goes in the direction of considering
models with conserved charges that can be studied in the large charge
limit~\cite{Hellerman:2015nra}.

In $d=d_c+\varepsilon$, for $\varepsilon>0,$ the fixed points that we have discovered
become RG unstable. A natural question that arises, then, is if there
exists another theory with the same global symmetries but, perhaps, defined
in a different dimension for which these same fixed points appear as stable
fixed points in the infrared. Similar considerations have been discussed in
\cite{Fei:2014yja} for the standard $O(N)$ model and could potentially be
extended to the theory \eqref{ONpot}.

\vskip 0.5cm
\noindent
{\bf Acknowledgements}
G.P.V.\ thanks Luca Zambelli for discussions.  Early stages of the
investigations presented in this work have been supported by the ACRI grant
``FRGIM'' within the YITP program.  Research presented in this article was
supported by the Laboratory Directed Research and Development program of
Los Alamos National Laboratory under project number 20180709PRD1. A.S.\ is
funded by the Royal Society under the grant ``Advancing the Conformal
Bootstrap Program in Three and Four Dimensions.''
\vskip -0.4cm
\appendix
\section{Useful Fourier transforms}
Our perturbative computations are based on combinations of Green functions and their derivatives which, depending on the structure of the Feynman diagrams,
are conveniently considered in either
coordinate or momentum space representation. Indeed, using an analogy with electric circuits, one can combine propagators attached to vertices in parallel or in series: melonic products of $2$-point functions (i.e.\ parallel combinations)
are more easily represented as products in coordinate space,
instead series combinations are more effectively studied as products of factors in momentum space.

We summarize below some useful relations involving the propagators, their spatial derivatives and their products.
\subsection{For the one loop beta function}\label{abeta}
The main building blocks in the computation are the Green function
$G(x) = \frac{c}{|x|^{2\delta}}$ and its derivatives
\be
\partial_\mu G_x = -\frac{2\delta c x^\mu}{|x|^{2\delta+2}}\,, \qquad
\partial_\mu\partial_\nu G_x = -\frac{2\delta c \delta^\mu_\nu}{|x|^{2\delta+2}} +\frac{2\delta(2\delta+2) c x^\mu x^\nu}{|x|^{2\delta+4}}\,,
\ee
and
\be  \label{ddDddG}
\partial_\mu\partial_\nu G_x \partial_\rho\partial_\sigma G_x = \frac{4\delta^2c^2\delta^\mu_\nu\delta^\rho_\sigma}{|x|^{4\delta+4}} - \frac{8\delta^2c^2(\delta+1)(x^\mu x^\nu\delta^\rho_\sigma+x^\rho
 x^\sigma \delta^\mu_\nu)}{|x|^{4\delta+6}} + \frac{16\delta^2c^2(\delta+1)^2 x^\mu x^\nu x^\rho x^\sigma}{|x|^{4\delta+8}}
\ee
Taking into account the fact that $\delta=\delta_c-\varepsilon/2$ and $\delta_c=k-2$, one finds the following Fourier transforms
{\setlength\arraycolsep{2pt}
\bea
\int_x \!e^{ip\cdot x} \;\frac{1}{|x|^{4\delta+4}} &=& \frac{2\pi^{2k-2}}{\Gamma(2k-2)\varepsilon} + \mathrm{finite} \
\eea}%
{\setlength\arraycolsep{2pt}
\bea
\int_x \!e^{ip\cdot x} \;\frac{x^\mu x^\nu}{|x|^{4\delta+6}}
&=& \frac{\pi^{2k-2}}{\Gamma(2k-1)\varepsilon}\,\delta^\mu_\nu  + \mathrm{finite}
\eea}%
{\setlength\arraycolsep{2pt}
\bea
\int_x \!e^{ip\cdot x} \;\frac{x^\mu x^\nu x^\rho x^\sigma}{|x|^{4\delta+6}}  &=& \frac{k\pi^{2k-2}}{\Gamma(2k+1)\varepsilon}\,(\delta^\mu_\nu \delta^\rho_\sigma+\delta^\mu_\rho \delta^\nu_\sigma + \delta^\mu_\sigma \delta^\nu_\rho) + \mathrm{finite}
\eea}%
{\setlength\arraycolsep{2pt}
\bea  \label{ddDddGdiv}
\int_x \!e^{ip\cdot x} \; \partial_\mu\partial_\nu G_x \partial_\rho\partial_\sigma G_x
&=& \frac{1}{2(4\pi)^{2k-2}\Gamma (2k)}\frac{1}{\varepsilon} (\delta^\mu_\nu \delta^\rho_\sigma + \delta^\mu_\rho \delta^\nu_\sigma +\delta^\mu_\sigma \delta^\nu_\rho)  + \mathrm{finite}
\eea}%
where some divergent contributions have canceled out. Some particular
contractions are needed in the main text
\be
\int_x \!e^{ip\cdot x} \; \partial_\rho\partial_\nu G_x \partial_\rho\partial_\sigma G_x = \frac{1}{2(4\pi)^{2k-2}\Gamma (2k)}\frac{1}{\varepsilon} (d_c+2)\delta^\nu_\sigma  + \mathrm{finite}\,,
\ee
\be
\int_x \!e^{ip\cdot x} \; \partial_\mu\partial_\nu G_x \partial_\mu\partial_\nu G_x = \frac{1}{2(4\pi)^{2k-2}\Gamma (2k)}\frac{1}{\varepsilon} d_c(d_c+2)  + \mathrm{finite}\,.
\ee
\subsection{For the two loop anomalous dimension}\label{aanom}
Starting from Eq.~\eqref{ddDddG}, one obtains
\be
(\partial_\mu\partial_\nu G)^2 = \frac{8\delta^2c^2(2\delta^2+3\delta+k)}{|x|^{4\delta+4}}
\ee
Here we list some useful expression related to triangle diagrams (with three propagators):
\be
(\partial_\rho\partial_\sigma G)^2 \partial_\mu\partial_\nu G = 16\delta^3c^3(2\delta^2+3\delta+k) \left[\frac{2(\delta+1)x_\mu x_\nu}{|x|^{6\delta+8}} -\frac{\delta_{\mu\nu}}{|x|^{6\delta+6}}\right]\,,
\ee
\be
\partial_\mu\partial_\sigma G\, \partial_\sigma\partial_\rho G = 4\delta^2 c^2 \left[\frac{\delta_{\mu\rho}}{|x|^{4\delta+4}} + \frac{4\delta(\delta+1)x_\mu x_\rho}{|x|^{4\delta+6}}\right]\,,
\ee
{\setlength\arraycolsep{2pt}
\bea
\partial_\mu\partial_\sigma G\, \partial_\sigma\partial_\rho G\, \partial_\rho\partial_\nu G &=& 8\delta^3 c^3 \left[\frac{2(\delta+1)(4\delta^2+2\delta+1)x_\mu x_\nu}{|x|^{6\delta+8}}  -\frac{\delta_{\mu\nu}}{|x|^{6\delta+6}}\right]\,.
\eea}%
Computing the following two Fourier transforms
{\setlength\arraycolsep{2pt}
\bea
\int_x \!e^{ip\cdot x} \;\frac{1}{|x|^{6\delta+6}} &=& \frac{(-1)^{k-1}4^{1-k}\pi^{2k-2}}{\Gamma(k)\Gamma(3k-3)\varepsilon} (p^2)^{k-1} + \mathrm{finite}\,, \nonumber
\eea}%
{\setlength\arraycolsep{2pt}
\bea
\int_x \!e^{ip\cdot x} \;\frac{x_\mu x_\nu}{|x|^{6\delta+8}}  &=& \frac{(-1)^{k-1}4^{1-k}\pi^{2k-2}}{\Gamma(k-1)\Gamma(3k-2)\varepsilon} (p^2)^{k-2} p_\mu p_\nu + \frac{(-1)^{k-1}2^{1-2k}\pi^{2k-2}}{\Gamma(k)\Gamma(3k-2)\varepsilon} (p^2)^{k-1}\delta_{\mu\nu}+ \mathrm{finite}\,, \nonumber
\eea}%
then plugging them into the previous expressions and returning to the coordinate representation, we get
\be
(\partial_\rho\partial_\sigma G)^2 \partial_\mu\partial_\nu G = \frac{(-1)^k (4\pi)^{4-4k}}{\Gamma (k) \Gamma (3 k-2)\varepsilon} (-\Box)^{k-1}\delta_{\mu\nu} - \frac{(-1)^{k+1}(4\pi)^{4-4k} }{\Gamma (k-1) \Gamma (3 k-2)\varepsilon} (-\Box)^{k-2} \partial_\mu \partial_\nu + \mathrm{finite} \,, \nonumber
\ee
{\setlength\arraycolsep{2pt}
\bea
\partial_\mu\partial_\sigma G\, \partial_\sigma\partial_\rho G\, \partial_\rho\partial_\nu G&=&  \frac{(-1)^{k+1} (4\pi)^{4-4k} (2 k-5) }
{4(k-1) \Gamma (k) \Gamma (3k-2)\varepsilon} (-\Box)^{k-1}\delta_{\mu\nu} \nn\\
&-& \frac{(-1)^{k+1}(4\pi)^{4-4k}(2 k (2 k-7)+13) }{4(k-1)\Gamma(k)\Gamma(3 k-2) \varepsilon}(-\Box)^{k-2} \partial_\mu \partial_\nu + \mathrm{finite} \,. \nonumber
\eea}%
\subsection{For the one loop counterterm of the cubic interaction}\label{betacubic}

We enumerate now the different terms which contribute to Eq.~\eqref{cubicsumCT}.
In the group $E_1$ there are $9$ different contributions, which belong to two different classes that are equivalent under permutations,
$6$ of them are equal to $E_{11}$, while $3$ are equal to $E_{17}$, according to the following definitions
\bea
E_{11}&:& \quad \int \prod_{i=1}^3 \dm^d x_i \partial_\mu\phi(x_1) \partial_\nu\phi(x_2) \Box\phi(x_3)
\partial_{1,\mu} \Box_2 G_{12} \partial_{2,\nu} \partial_{3,\sigma} G_{23} \partial_{3,\sigma} \Box_1 G_{31}
\nonumber \\
  E_{17}&:& \quad  \int \prod_{i=1}^3 \dm^d x_i \partial_\mu\phi(x_1) \partial_\nu\phi(x_2) \Box\phi(x_3)
\Box_1 \partial_{2,\nu} G_{12} \Box_2 \partial_{3,\sigma} G_{23} \partial_{3,\sigma} \partial_{1,\mu} G_{31} \,.
 \eea
These correspond to two one loop triangle diagrams and can be evaluated using the Fourier transforms
$\phi(x_i)=\int \frac{\dm^d p_i}{(2\pi)^d} \tilde \phi(p_i) e^{-i p_i x_i}$ and
$G_{ij}=\int \frac{\dm^d q_l}{(2\pi)^d} \frac{1}{(q_l^2)^k} e^{i q_l (x_i-x_j)}$,
and introducing the momenta $k_i$ such that $p_i=k_{i+1}-k_i$ in cyclic form. Performing a Feynman parametrization
with parameters $a_i\,,~i=1,2,3$, so that one has integrals of the form
\be
\int_0^1 \prod_i \dm a_i \delta(1-\sum_i a_i) \prod_i a_i^{\alpha_i-1} \int \frac{\dm^d q}{(2\pi)^d} \frac{N(q,k_i)}{\left[ (q+\sum_i a_i k_i)^2+\Delta\right]^{\sum_i\alpha_i}}\,,
\ee
with suitable $\alpha_i$ and $\Delta=\sum_{i<j} a_i a_j k_{ij}^2=a_1a_2 p_1^2+a_2a_3 p_2^2+ a_2a_1 p_3^2$, with $k_{ij}=k_i-k_j$.
The $q$ integration on the above terms leads to the expected UV divergent term with the $1/\varepsilon$ pole.
In order to compute the intergrals in the Feynman parameters, it is convenient to use the Mellin-Barnes parametrization to disantangle the power of the sum present in $\Delta$, according to the formula
\be
\frac{1}{\Delta^\gamma}=\int \frac{\dm y \dm z}{(2\pi i)^2} (a_1 a_2 p_1^2)^{-\gamma-y-z} (a_2 a_3 p_2^2)^y (a_3 a_1 p_3^2)^z
\frac{\Gamma(-y) \Gamma(-z) \Gamma(\gamma+y+z)}{\Gamma(\gamma)} \,,
\ee
where the contours of the integrations are chosen such that the poles of $\Gamma(-y)$ and $\Gamma(-z)$ are to their right and the poles in $\Gamma(\gamma+y+z)$ are to their left.
Using all the above methods, we find
\bea
(E_{1i})_{div}&=&\frac{2}{(4\pi)^{3k-4} \Gamma(3k-3)} \frac{1}{\varepsilon} \int \dm^d x \left(\frac{1}{2} \partial \phi \cdot \partial \phi \Box \phi\right)
\quad i=1,\cdots ,6 \nonumber \\
(E_{1i})_{div}&=&- (E_{11})_{div} \quad i=7,8, 9 \,.
\eea
Then there are three equal contributions $E_{2i}$ with $i,1,2,3$ of the form
\be
E_{21}: \quad \int \prod_{i=1}^3 \dm^d x_i \partial_\mu\phi(x_1) \Box \phi(x_2) \partial_\sigma \phi(x_3)
\partial_{1,\mu} \partial_{2,\nu} G_{12} \partial_{2,\nu} \partial_{3,\sigma} G_{23} \Box_3 \Box_1 G_{31}
\ee
plus permutations,
which give
\be
(E_{2i})_{div}=- (E_{11})_{div} \quad i=1,2, 3 \,.
\ee
We then have six contributions denoted $E_{3i}$, which divide into two classes of three
 \bea
E_{31}&:& \quad \int \prod_{i=1}^3 \dm^d x_i \partial_\mu\phi(x_1) \partial_\nu\phi(x_2) \partial_\sigma\phi(x_3)
\partial_{1,\mu} \Box_2 G_{12} \partial_{2,\nu} \partial_{3,\sigma} G_{23} \Box_3 \Box_1 G_{31}
\nonumber \\
  E_{34}&:& \quad  \int \prod_{i=1}^3 \dm^d x_i \partial_\mu\phi(x_1) \partial_\nu\phi(x_2) \partial_\sigma\phi(x_3)
\Box_1 \partial_{2,\nu} G_{12} \Box_2  \Box_3 G_{23} \partial_{3,\sigma} \partial_{1,\mu} G_{31} \,.
 \eea
After an explicit computation we find
 \be
(E_{3i})_{div}=-\frac{1}{(4\pi)^{3k-4} \Gamma(3k-3)} \frac{1}{\varepsilon} \int \dm^d x \left(\frac{1}{2} \partial \phi \cdot \partial \phi \Box \phi\right)
\quad i=1, \cdots, 6
\ee
There are also two contributions denoted by $E_{4i}$, $i=1,2$ which are all equal. The first is defined by
\be
E_{41}: \quad \int \prod_{i=1}^3 \dm^d x_i \partial_\mu\phi(x_1) \Box \phi(x_2) \partial_\sigma \phi(x_3)
\partial_{1,\mu} \Box_2 G_{12} \partial_{2,\nu} \partial_{3,\sigma} G_{23} \Box_3 \Box_1 G_{31} \,.
\ee
We find
\be
(E_{4i})_{div}=\frac{3}{(4\pi)^{3k-4} \Gamma(3k-3)} \frac{1}{\varepsilon} \int \dm^d x \left(\frac{1}{2} \partial \phi \cdot \partial \phi \Box \phi\right)
\quad i=1, 2\,.
\ee
Finally there are also six terms of the form
\be
E_{51}: \quad \int \prod_{i=1}^3 \dm^d x_i \Box \phi(x_1) \Box \phi(x_2) \partial_\sigma \phi(x_3)
\partial_{1,\mu} \partial_{2,\nu} G_{12} \partial_{2,\nu} \partial_{3,\sigma} G_{23} \Box_3 \partial_{1,\mu}  G_{31}
\ee
and a last one
\be
E_{6}: \quad \int \prod_{i=1}^3 \dm^d x_i \Box \phi(x_1) \Box \phi(x_2) \Box \phi(x_3)
\partial_{1,\mu} \partial_{2,\nu} G_{12} \partial_{2,\nu} \partial_{3,\sigma} G_{23}  \partial_{3,\sigma}  \partial_{1,\mu}  G_{31} \,,
\ee
which are finite (they have no UV divergences) and therefore do not contribute the perturbative renormalization of the coupling.
\bibliographystyle{chetref}
\bibliography{biblio}

\begin{thebibliography}{10}
\ifx\href\asklfhas\newcommand{\href}[2]{#2}\fi
\ifx\arxivref\asklfhas\newcommand{\arxivref}[2]{\href{http://arxiv.org/abs/#1}{#2}}\fi
\ifx\doiref\asklfhas\newcommand{\doiref}[2]{\href{http://dx.doi.org/#1}{#2}}\fi
\parskip 0pt
\normalsize

\bibitem{Fisher:1978pf}
M.~E. Fisher,
\textit{``{Yang-Lee Edge Singularity and $\phi^3$ Field Theory}''},
\doiref{10.1103/PhysRevLett.40.1610}{Phys.~Rev.~Lett. \textbf{40}, 1610
  (1978)\ignorespaces}\ignorespaces
\bibitem{Cardy:1985yy}
J.~L. Cardy,
\textit{``{Conformal Invariance and the Yang-Lee Edge Singularity in
  Two-dimensions}''},
\doiref{10.1103/PhysRevLett.54.1354}{Phys.~Rev.~Lett. \textbf{54}, 1354
  (1985)\ignorespaces}\ignorespaces
\bibitem{Landau:1986aog}
L.~D. Landau \& E.~M. Lifshitz,
\textit{``{Theory of Elasticity}''},
Elsevier Butterworth-Heinemann (1986)\ignorespaces,
New York
\bibitem{Riva:2005gd}
V.~Riva \& J.~L. Cardy,
\textit{``{Scale and conformal invariance in field theory: A Physical
  counterexample}''},
\doiref{10.1016/j.physletb.2005.07.010}{Phys.~Lett.~B \textbf{622}, 339
  (2005)\ignorespaces}\ignorespaces,
\normalsize{\texttt{\arxivref{hep-th/0504197}{hep-th/0504197}}}\ignorespaces
\bibitem{David:1992vv}
F.~David, B.~Duplantier \& E.~Guitter,
\textit{``{Renormalization theory for interacting crumpled manifolds}''},
\doiref{10.1016/0550-3213(93)90226-F}{Nucl.~Phys.~B \textbf{394}, 555
  (1993)\ignorespaces}\ignorespaces,
\normalsize{\texttt{\arxivref{hep-th/9211038}{hep-th/9211038}}}\ignorespaces
\bibitem{David:1993rr}
F.~David, B.~Duplantier \& E.~Guitter,
\textit{``{Renormalization of crumpled manifolds}''},
\doiref{10.1103/PhysRevLett.70.2205}{Phys.~Rev.~Lett. \textbf{70}, 2205
  (1993)\ignorespaces}\ignorespaces,
\normalsize{\texttt{\arxivref{hep-th/9212102}{hep-th/9212102}}}\ignorespaces
\bibitem{Polchinski:1987dy}
J.~Polchinski,
\textit{``{Scale and Conformal Invariance in Quantum Field Theory}''},
\doiref{10.1016/0550-3213(88)90179-4}{Nucl.~Phys.~B \textbf{303}, 226
  (1988)\ignorespaces}\ignorespaces
\bibitem{Nakayama:2013is}
Y.~Nakayama,
\textit{``{Scale invariance vs conformal invariance}''},
\doiref{10.1016/j.physrep.2014.12.003}{Phys.~Rept. \textbf{569}, 1
  (2015)\ignorespaces}\ignorespaces,
\normalsize{\texttt{\arxivref{1302.0884}{arXiv:1302.0884
  \![hep-th]}}}\ignorespaces
\bibitem{Wetterich:1992yh}
C.~Wetterich,
\textit{``{Exact evolution equation for the effective potential}''},
\doiref{10.1016/0370-2693(93)90726-X}{Phys.~Lett.~B \textbf{301}, 90
  (1993)\ignorespaces}\ignorespaces,
\normalsize{\texttt{\arxivref{1710.05815}{arXiv:1710.05815
  \![hep-th]}}}\ignorespaces
\bibitem{Morris:1994ie}
T.~R. Morris,
\textit{``{Derivative expansion of the exact renormalization group}''},
\doiref{10.1016/0370-2693(94)90767-6}{Phys.~Lett.~B \textbf{329}, 241
  (1994)\ignorespaces}\ignorespaces,
\normalsize{\texttt{\arxivref{hep-ph/9403340}{hep-ph/9403340}}}\ignorespaces
\bibitem{Safari:2017irw}
M.~Safari \& G.~P. Vacca,
\textit{``{Multicritical scalar theories with higher-derivative kinetic terms:
  A perturbative RG approach with the $\varepsilon$-expansion}''},
\doiref{10.1103/PhysRevD.97.041701}{Phys.~Rev.~D \textbf{97}, 041701
  (2018)\ignorespaces}\ignorespaces,
\normalsize{\texttt{\arxivref{1708.09795}{arXiv:1708.09795
  \![hep-th]}}}\ignorespaces
\bibitem{Safari:2017tgs}
M.~Safari \& G.~P. Vacca,
\textit{``{Uncovering novel phase structures in $\Box ^k$ scalar theories with
  the renormalization group}''},
\doiref{10.1140/epjc/s10052-018-5721-4}{Eur.~Phys.~J.~C \textbf{78}, 251
  (2018)\ignorespaces}\ignorespaces,
\normalsize{\texttt{\arxivref{1711.08685}{arXiv:1711.08685
  \![hep-th]}}}\ignorespaces
\bibitem{Codello:2017hhh}
A.~Codello, M.~Safari, G.~P. Vacca \& O.~Zanusso,
\textit{``{Functional perturbative RG and CFT data in the
  $\varepsilon$-expansion}''},
\doiref{10.1140/epjc/s10052-017-5505-2}{Eur.~Phys.~J.~C \textbf{78}, 30
  (2018)\ignorespaces}\ignorespaces,
\normalsize{\texttt{\arxivref{1705.05558}{arXiv:1705.05558
  \![hep-th]}}}\ignorespaces
\bibitem{Rychkov:2015naa}
S.~Rychkov \& Z.~M. Tan,
\textit{``{The $\epsilon$-expansion from conformal field theory}''},
\doiref{10.1088/1751-8113/48/29/29FT01}{J.~Phys.~A \textbf{48}, 29FT01
  (2015)\ignorespaces}\ignorespaces,
\normalsize{\texttt{\arxivref{1505.00963}{arXiv:1505.00963
  \![hep-th]}}}\ignorespaces
\bibitem{Codello:2017qek}
A.~Codello, M.~Safari, G.~P. Vacca \& O.~Zanusso,
\textit{``{Leading CFT constraints on multi-critical models in d \ensuremath{>}
  2}''},
\doiref{10.1007/JHEP04(2017)127}{JHEP \textbf{1704}, 127
  (2017)\ignorespaces}\ignorespaces,
\normalsize{\texttt{\arxivref{1703.04830}{arXiv:1703.04830
  \![hep-th]}}}\ignorespaces
\bibitem{Osborn:2016bev}
H.~Osborn \& A.~Stergiou,
\textit{``{C$_{T}$ for non-unitary CFTs in higher dimensions}''},
\doiref{10.1007/JHEP06(2016)079}{JHEP \textbf{1606}, 079
  (2016)\ignorespaces}\ignorespaces,
\normalsize{\texttt{\arxivref{1603.07307}{arXiv:1603.07307
  \![hep-th]}}}\ignorespaces
\bibitem{OsbornLectures}
H.~Osborn,
\textit{``{Lectures on Conformal Field Theories in more than two
  dimensions}''},
\href{http://www.damtp.cam.ac.uk/user/ho/CFTNotes.pdf}{\texttt{http://www.damtp.cam.ac.uk/user/ho/CFTNotes.pdf}}
\bibitem{Guerrieri:2016whh}
A.~Guerrieri, A.~C. Petkou \& C.~Wen,
\textit{``{The free $\sigma$CFTs}''},
\doiref{10.1007/JHEP09(2016)019}{JHEP \textbf{1609}, 019
  (2016)\ignorespaces}\ignorespaces,
\normalsize{\texttt{\arxivref{1604.07310}{arXiv:1604.07310
  \![hep-th]}}}\ignorespaces
\bibitem{Romoli:2021hre}
M.~Romoli \& O.~Zanusso,
\textit{``{A different kind of four dimensional brane for string theory}''},
\normalsize{\texttt{\arxivref{2110.05584}{arXiv:2110.05584
  \![hep-th]}}}\ignorespaces
\bibitem{Coquand:2020tgb}
O.~Coquand, D.~Mouhanna \& S.~Teber,
\textit{``{Flat phase of polymerized membranes at two-loop order}''},
\doiref{10.1103/PhysRevE.101.062104}{Phys.~Rev.~E \textbf{101}, 062104
  (2020)\ignorespaces}\ignorespaces,
\normalsize{\texttt{\arxivref{2003.13973}{arXiv:2003.13973
  \![cond-mat.stat-mech]}}}\ignorespaces
\bibitem{Metayer:2021kxm}
S.~Metayer, D.~Mouhanna \& S.~Teber,
\textit{``{A three-loop order approach to flat polymerized membranes}''},
\normalsize{\texttt{\arxivref{2109.03796}{arXiv:2109.03796
  \![cond-mat.stat-mech]}}}\ignorespaces
\bibitem{Osborn:2017ucf}
H.~Osborn \& A.~Stergiou,
\textit{``{Seeking fixed points in multiple coupling scalar theories in the
  $\varepsilon$ expansion}''},
\doiref{10.1007/JHEP05(2018)051}{JHEP \textbf{1805}, 051
  (2018)\ignorespaces}\ignorespaces,
\normalsize{\texttt{\arxivref{1707.06165}{arXiv:1707.06165
  \![hep-th]}}}\ignorespaces
\bibitem{Codello:2019isr}
A.~Codello, M.~Safari, G.~P. Vacca \& O.~Zanusso,
\textit{``{Symmetry and universality of multifield interactions in
  $6-\varepsilon$ dimensions}''},
\doiref{10.1103/PhysRevD.101.065002}{Phys.~Rev.~D \textbf{101}, 065002
  (2020)\ignorespaces}\ignorespaces,
\normalsize{\texttt{\arxivref{1910.10009}{arXiv:1910.10009
  \![hep-th]}}}\ignorespaces
\bibitem{Codello:2020lta}
A.~Codello, M.~Safari, G.~P. Vacca \& O.~Zanusso,
\textit{``{Critical models with $N \leq $4 scalars in $d=4-\varepsilon$}''},
\doiref{10.1103/PhysRevD.102.065017}{Phys.~Rev.~D \textbf{102}, 065017
  (2020)\ignorespaces}\ignorespaces,
\normalsize{\texttt{\arxivref{2008.04077}{arXiv:2008.04077
  \![hep-th]}}}\ignorespaces
\bibitem{Osborn:2020cnf}
H.~Osborn \& A.~Stergiou,
\textit{``{Heavy handed quest for fixed points in multiple coupling scalar
  theories in the $\varepsilon$ expansion}''},
\doiref{10.1007/JHEP04(2021)128}{JHEP \textbf{2104}, 128
  (2021)\ignorespaces}\ignorespaces,
\normalsize{\texttt{\arxivref{2010.15915}{arXiv:2010.15915
  \![hep-th]}}}\ignorespaces
\bibitem{Codello:2018nbe}
A.~Codello, M.~Safari, G.~P. Vacca \& O.~Zanusso,
\textit{``{Leading order CFT analysis of multi-scalar theories in
  d\ensuremath{>}2}''},
\doiref{10.1140/epjc/s10052-019-6817-1}{Eur.~Phys.~J.~C \textbf{79}, 331
  (2019)\ignorespaces}\ignorespaces,
\normalsize{\texttt{\arxivref{1809.05071}{arXiv:1809.05071
  \![hep-th]}}}\ignorespaces
\bibitem{Gracey:2015xmw}
J.~A. Gracey,
\textit{``{Six dimensional QCD at two loops}''},
\doiref{10.1103/PhysRevD.93.025025}{Phys.~Rev.~D \textbf{93}, 025025
  (2016)\ignorespaces}\ignorespaces,
\normalsize{\texttt{\arxivref{1512.04443}{arXiv:1512.04443
  \![hep-th]}}}\ignorespaces
\bibitem{Gracey:2017erc}
J.~A. Gracey \& R.~M. Simms,
\textit{``{Higher dimensional higher derivative $\phi^4$ theory}''},
\doiref{10.1103/PhysRevD.96.025022}{Phys.~Rev.~D \textbf{96}, 025022
  (2017)\ignorespaces}\ignorespaces,
\normalsize{\texttt{\arxivref{1705.06983}{arXiv:1705.06983
  \![hep-th]}}}\ignorespaces
\bibitem{Hellerman:2015nra}
S.~Hellerman, D.~Orlando, S.~Reffert \& M.~Watanabe,
\textit{``{On the CFT Operator Spectrum at Large Global Charge}''},
\doiref{10.1007/JHEP12(2015)071}{JHEP \textbf{1512}, 071
  (2015)\ignorespaces}\ignorespaces,
\normalsize{\texttt{\arxivref{1505.01537}{arXiv:1505.01537
  \![hep-th]}}}\ignorespaces
\bibitem{Fei:2014yja}
L.~Fei, S.~Giombi \& I.~R. Klebanov,
\textit{``{Critical $O(N)$ models in $6-\epsilon$ dimensions}''},
\doiref{10.1103/PhysRevD.90.025018}{Phys.~Rev.~D \textbf{90}, 025018
  (2014)\ignorespaces}\ignorespaces,
\normalsize{\texttt{\arxivref{1404.1094}{arXiv:1404.1094
  \![hep-th]}}}\ignorespaces
\end{thebibliography}

\end{document}